\documentclass[epj]{svjour}
\usepackage{graphicx}
\usepackage{amsmath}
\usepackage{amssymb}
\begin{document}
%
%\title{Physisorption kinetics of surface charges at plasma boundaries}
\title{Physisorption kinetics of electrons at plasma boundaries}
%\author{F. X. Bronold \and H. Fehske \and H. Deutsch %etc
\author{F. X. Bronold \and H. Deutsch \and H. Fehske
}                     % Do not remove
\institute{Institut f\"ur Physik, Ernst-Moritz-Arndt-Universit\"at Greifswald, D-17489 Greifswald, Germany}
\date{Received: date / Revised version: date}
\abstract{
Plasma-boundaries floating in an ionized gas are usually negatively charged. They 
accumulate electrons more efficiently than ions leading to the formation of a 
quasi-stationary electron film at the boundaries. We propose to interpret the build-up
of surface charges at inert plasma boundaries, where other surface modifications, 
for instance, implantation of particles and reconstruction or destruction of the 
surface due to impact of high energy particles can be neglected, as a physisorption
process in front of the wall. The electron sticking coefficient $s_e$ and the electron 
desorption time $\tau_e$, which play an important role in determining the 
quasi-stationary surface charge, and about which little is empirically and theoretically 
known, can then be calculated from microscopic models for the electron-wall interaction.
Irrespective of the sophistication of the models, the static part of the electron-wall
interaction determines the binding energy of the electron, whereas inelastic processes 
at the wall determine $s_e$ and $\tau_e$. As an illustration, we calculate $s_e$ and $\tau_e$ 
for a metal, using the simplest model in which the static part of the electron-metal 
interaction is approximated by the classical image potential. Assuming electrons from the 
plasma to loose (gain) energy at the surface by creating (annihilating) electron-hole 
pairs in the metal, which is treated as a jellium half-space with an infinitely high 
workfunction, we obtain $s_e\approx 10^{-4}$ and $\tau_e\approx 10^{-2}s$. The product 
$s_e\tau_e\approx 10^{-6}s$ has the order of magnitude expected from our earlier 
results for the charge of dust particles in a plasma but individually $s_e$ is 
unexpectedly small and $\tau_e$ is somewhat large. The former is a consequence of 
the small matrix elements occurring in the simple model while the latter is due to 
the large binding energy of the electron. More sophisticated theoretical investigations, 
but also experimental support, are clearly needed because if $s_e$ is indeed as small as our 
exploratory calculation suggests, it would have severe consequences for the understanding 
of the formation of surface charges at plasma boundaries. To identify what we believe are
key issues of the electronic microphysics at inert plasma boundaries and to inspire other 
groups to join us on our journey is the purpose of this colloquial presentation.
\PACS{
      {52.27.Lw}{Dusty or complex plasmas}\and
      {52.40.Hf}{Plasma-material interaction, boundary layer effects}\and
      {68.43.-h}{Chemi-/Physisorption: adsorbates on surfaces}\and
      {73.20.-r}{Electron states at surfaces and interfaces}
     } % end of PACS codes
} %end of abstract
\maketitle
\section{Introduction}
\label{intro}

Low-temperature plasma physics is undoubtedly an applied science driven by the
ever increasing demand for plasma-assisted surface modification processes and
environmentally save, low-power consuming lighting devices. At the same time,
however, the physics of gas discharges is rich on fundamental problems which are
of broader interest.

From a formal point of view, a gas discharge is an externally driven bounded
reactive multicomponent system. It contains, besides electrons and ions,
chemically reactive atoms and/or molecules strongly interacting with each other
and with external (wall of the discharge vessel) as well as internal ($nm$ to
$\mu m$-sized solid particles) boundaries. Like in any reactive system elementary
collision processes (elastic, inelastic, and reactive), occurring on a
microscopic scale, determine in conjunction with external control parameters the
global properties of the system on the macroscopic scale. However, whereas in an 
ordinary chemical reactor all constituents are neutral, a gas discharge contains
also charged constituents. There are thus at least two macroscopic scales: 
the electromagnetic scale, where screening and sheath formation takes 
place~\cite{Franklin06,Riemann91}, and the extension of the vessel. Since the observed 
physical properties of a gas discharges emerge from processes occurring on at least three 
different length (and time) scales -- one microscopic and two macroscopic scales -- 
the starting point of any quantitative description 
is a multiple-scale analysis even if it is not explicitly performed. Being externally 
driven, low-temperature plasmas are moreover far-off thermal equilibrium and like other
dissipative systems feature a great variety of self-organization 
phenomena~\cite{Purwins08,PBL04}. Finally, and this sets the theme of this colloquium,
low-temperature gas discharges, in contrast to magnetically confined 
high-temperature fusion plasmas, are directly bounded by massive macroscopic objects. 
Thus, they strongly interact with solids.

The plasma-solid interaction is of course at the core of all plasma-assisted
surface processes (deposition, implantation, sputtering, etching, etc.)~\cite{LL05}.
Of more fundamental interest, however, is the situation of a chemically inert (i.e., 
no surface modification due to chemical processes, no reconstruction or destruction 
of the surface due to high-energy particles etc.) floating 
surface, where the interaction with the plasma leads only to the build-up of surface
charges and thus to a quasi-two-dimensional electron film which may have unique 
properties similar to electrons trapped on a liquid helium surface~\cite{Cole74} or 
to electrons confined in a semiconductor heterojunction~\cite{AFS82}.

In plasma-physical settings surface charges play a role in atmospheric 
plasmas, where the charge of $nm$-sized aerosols~\cite{RL01} is of interest, in 
space bound plasmas, where surface charges of spacecrafts~\cite{GW00,Whipple81}
and of interplanetary and interstellar dust particles~\cite{Mann08,Horanyi96} have 
been extensively studied, and in laboratory dusty plasmas, where the study of 
self-organization of highly negatively charged, strongly interacting $\mu m$-sized 
dust particles became an extremely active area of current plasma 
research~\cite{Ishihara07,FIK05,KRZ05,SV03,TAA00,TLA00,WHR95}.   
Surface charges affect also the physics of dielectric barrier discharges -- a 
discharge type of huge technological impact~\cite{GMB02,Kogelschatz03,KCO04,SAB06,SLP07,LLZ08}.

That surface charges at plasma boundaries could be considered as a thin film of 
adsorbed electrons (\textquotedblleft surface plasma\textquotedblright) in contact 
with the bulk plasma was originally suggested by Emeleus and Coulter in connection
with their investigations of wall recombination in the positive column~\cite{EC87}. 
Later, Behnke and coworkers~\cite{BBD97} used this idea to phenomenologically construct 
boundary conditions for the kinetic equations describing glow discharges and Kersten 
{\it et al.}~\cite{KDK04} employed the notion of a surface plasma to study the charging 
of dust particles in a plasma.

Although the surface plasma as a physical entity with its own physical properties is 
implicitly contained in these investigations, a microscopic description of its formation, 
dynamics, and structure was not attempted. First steps in this direction were taken by us 
in a short note~\cite{BFK08}. The purpose of this colloquium is, on the one hand, to extend 
these considerations, in particular, to identify the surface physics 
which needs to be resolved before a quantitative microscopic 
theory of the surface plasma can be constructed and to convey, on the other hand, our 
conviction that the concept itself is not empty. On the contrary, it puts questions 
center stage which are of fundamental interest. To list just a few:
\begin{center}
\begin{tabbing}
$\bullet$ \= What forces bind electrons and ions to the plasma\\
          \> boundary?\\
$\bullet$ \> How do electrons and ions dissipate energy when\\
          \> approaching the boundary?\\ 
$\bullet$ \> What is the probability with which an electron sticks\\
          \>  at or desorbs from the boundary?\\
$\bullet$ \> What is the density and temperature of the surface\\
          \> plasma and are there any collective properties?\\
$\bullet$ \> What is the mobility for the lateral motion of \\
          \> electrons and ions along the wall and can it be \\
          \> externally controlled?\\
$\bullet$ \> How does all this affect electron-ion recombination\\
          \> and secondary electron emission on chemically inert\\
          \> plasma boundaries?
\end{tabbing}
\end{center}

\begin{figure}[t]
\centering
\includegraphics[width=0.96\linewidth]{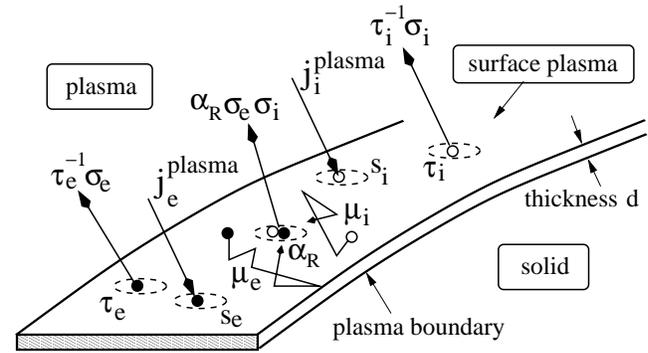}
\caption{Illustration of the elementary surface processes leading
to the build-up of a quasi-stationary surface plasma at an 
inert plasma boundary.} 
\label{Eprocesses}
\end{figure}

The elementary processes responsible for the formation of a surface plasma at an inert plasma
boundary are shown in Fig~\ref{Eprocesses}. Electrons and ions are collected 
from the plasma with collection fluxes $j_{e,i}^{\rm coll}=s_{e,i}j_{e,i}^{\rm plasma}$, 
where $s_{e,i}$ are the sticking coefficients and $j_{e,i}^{\rm plasma}$ are the fluxes of 
plasma electrons and ions hitting the boundary. Electrons and ions may thermally desorb from 
the boundary with rates $\tau_{e,i}^{-1}$, where $\tau_{e,i}$ are the desorption times. They 
may also move along the surface with mobilities $\mu_{e,i}$, which in turn may affect the 
probability $\alpha_R$ with which ions recombine with electrons at the wall. All these 
processes occur in a layer whose thickness $d$ is at most a few microns, that is, on a scale
where the standard kinetic description of the gas discharge based on the Boltzmann-Poisson 
system breaks down. Thus, the above listed questions can be only addressed from a 
quantum-mechanical point of view.

Of particular importance for the quantitative description of the build-up of a surface 
plasma are the sticking coefficients $s_{e,i}$ and the desorption times $\tau_{e,i}$. 
Little is quantitatively known about these parameters, in particular, with respect to
the electrons. Very often, $s_e\approx s_i\approx 0.1-1$ and $\tau_e^{-1}=\tau_i^{-1}=0$
is used without further justification. Below, we sketch a quantum-kinetic approach to 
calculate $s_e$ and $\tau_e$ from a simple microscopic model for the plasma boundary 
interaction which treats the interaction of electrons with plasma boundaries as a 
physisorption process~\cite{LJD36,BY73,GKT80a,GKT80b,KT81,Brenig82,KG86,NNS86,GS91,BR92,WJS92} 
in the polarization-induced attractive part of the surface potential. Electron surface 
states~\cite{RM72,EM73,Barton81,DAG84,SH84,WHJ85,JDK86,EP90,EL94,Fauster94,EL95,HSR97,CSE99,Hoefer99,VPE07}, 
at most a few $nm$ away from the boundary, will thus play a central role as will surface-bound
scattering processes which control electron energy relaxation at the surface and thus 
electron sticking and desorption. 

Although the forces and scales are different for ions, they behave
conceptually very similar. The main difference between electrons and ions is that 
as soon as the surface collected some electrons, because of the faster bombardment with 
electrons than with ions, the surface potential for ions is the attractive Coulomb 
potential (most probably screened but thats for the following irrelevant). Hence,  
ion surface states develop in the tail of the long-ranged Coulomb potential and thus 
deep in the sheath of the grain, far away from its surface. The 
microscopic processes driving ion energy relaxation and eventually ion sticking 
and desorption are thus not surface- but plasma-bound.

In the microscopic approach presented below, we focus on the 
physics occurring at most a few $nm$ away from the boundary. We will therefore
not give here a quantitative treatment of the physisorption kinetics of ions in 
the long-ranged Coulomb potential. However, when it comes to the calculation of 
the surface charge via phenomenological equations connecting the quantum with the 
classical level, we have to make some assumptions about the ion dynamics and 
kinetics. We will then discuss ions qualitatively. The assumptions made for ions,
which are somewhat in conflict with what other people expect~\cite{LGG01,LGS03,SLR04},
do however not affect the microscopic calculation of $s_e$ and $\tau_e$.

The outline of this colloquium is as follows. In the next section we describe 
and put into context the surface model for the charge of a floating dust particle 
in a plasma we developed in~\cite{BFK08} because it motivated the physisorption-inspired 
microscopic treatment of electrons at plasma boundaries discussed in this colloquium. 
A qualitative description of the ion kinetics in the vicinity of a spherical grain is 
also included in this section. Section 3 describes a microscopic model for the interaction 
of electrons with plasma boundaries. Specified to a metallic boundary, it will then be 
used to calculate the electron sticking coefficient $s_e$ and the electron desorption 
time $\tau_e$. Key issues of the microscopic description of the electron-wall interaction
(surface potential, coupling to elementary excitations of the solid, etc.) will be 
identified and numerical results will be presented and discussed. A critique of our
assumptions is given at the end of section 3 and should be understood as a list of 
to-do's. We close the presentation in section 4 with a few concluding remarks. 
Mathematical details interrupting the presentation which is meant to be read in order 
because it successively constructs a case are relegated to three appendices.

\section{Charge of a dust particle in a plasma}

The physisorption-inspired treatment of surface charges originated from our attempt
to calculate the charge of a spherical $\mu m$-sized floating dust particle in a 
quiescent plasma, taking not only plasma-induced but also surface-induced processes 
into account~\cite{BFK08}. Here we have to clearly distinguish between the assumptions
made to construct a constituting equation for the surface charge, which by necessity
has to connect the quantum mechanics occurring at the surface with the classical physics
determining the plasma fluxes, and the assumptions to obtain estimates for the surface 
parameters appearing in this equation. The microscopic calculation of the electron 
surface parameters $s_e$ and $\tau_e$ presented in the next sections is of course 
independent of the assumptions about the ion dynamics and kinetics as well as the 
phenomenological nature of the constituting equation for the surface charge.

\subsection{Rate equations}

First, we will discuss the surface model proposed in~\cite{BFK08} from the perspective 
of the rate equations corresponding to the elementary processes shown in 
Fig.~\ref{Eprocesses}. Thereby we also identify the assumptions, in particular, with 
respect to the surface properties, which are usually made in standard calculations of 
surface charges.

To be specific let us consider a spherical dust particle with radius $R$. The 
quasi-stationary charge of the grain is given by (we measure charge in units of $-e$)
\begin{eqnarray}
Z_p=4\pi R^2\big[\sigma_{e}-\sigma_i\big]~,
\end{eqnarray}
with electron and ion surface densities, $\sigma_{e,i}$, satisfying the
quasi-stationary ($d\sigma_{e,i}/dt=0$) rate equations~\cite{KDK04},
\begin{eqnarray}
0 &=& s_e j^{\rm plasma}_e-\tau_e^{-1}\sigma_e-\alpha_R\sigma_e\sigma_i~,
\label{REqe}\\
0 &=& s_i j^{\rm plasma}_i-\tau_i^{-1}\sigma_i-\alpha_R\sigma_e\sigma_i~,
\label{REqi}
\end{eqnarray}
where $j^{\rm plasma}_{e,i}$, $s_{e,i}$, $\tau_{e,i}$, and $\alpha_R$ denote, respectively, 
the fluxes of electrons and ions hitting the grain surface from the plasma, the electron and 
ion sticking coefficients, the electron and ion desorption times, and the electron-ion 
recombination coefficient.~\footnote{The rate equations connecting the
plasma fluxes $j_{e,i}^{\rm plasma}$ and surface densities $\sigma_{e,i}$ with the surface
parameters $s_{e,i}$, $\tau_{e,i}$, and $\alpha_R$ are phenomenological. They should be 
derived from Boltzmann equations containing surface scattering integrals which encapsulate 
the quantum mechanics responsible for sticking, desorption, and recombination.}

In order to derive the standard criterion invoked to determine the quasi-stationary grain 
charge, we now assume, in contrast to what we do in our model~\cite{BFK08} (see also below), 
that both electrons and ions reach the surface of the grain. In that case, both Eq.~(\ref{REqe}) 
and Eq.~(\ref{REqi}) should be interpreted as flux balances on the grain 
surface. At quasi-stationarity, the grain is charged to the floating potential $\bar{U}$. 
In energy units, $\bar{U}=Z_p e^2/R=2Z_pR_0a_B/R$ with $R_0$ the Rydberg energy and 
$a_B$ the Bohr radius. Because the grain temperature $k_BT_s\ll\bar{U}$ 
the ion desorption rate $\tau_i^{-1}\approx 0$. Equation~(\ref{REqi}) reduces 
therefore to $\alpha_R \sigma_e \sigma_i=s_i j^{\rm plasma}_i$ which transforms 
Eq.~(\ref{REqe}) into $s_e j^{\rm plasma}_e=s_ij^{\rm plasma}_i+\tau_e^{-1}\sigma$ provided 
$\sigma\approx\sigma_e$ which is usually the case. In the standard approach the grain surface 
is moreover assumed to be a perfect absorber for both electrons and ions. Thus, $s_e=s_i=1$
and $\tau^{-1}_e=\tau_i^{-1}=0$. The quasi-stationary charge $Z_p$ of the grain is then obtained
from the condition
\begin{eqnarray}
j^{\rm plasma}_e(Z_p)=j^{\rm plasma}_i(Z_p)~,
\label{OrdinaryApp}
\end{eqnarray}
where we explicitly indicated the dependence of the plasma fluxes on the grain charge. 

Calculations of the grain charge differ primarily in the approximations made for the plasma 
fluxes $j_{e,i}^{\rm plasma}$. For the repelled species, usually collisionless electrons, the flux
can be obtained from Poisson's equation and the collisionless Boltzmann equation, using trajectory 
tracing techniques based on Liouville's theorem and energy and momentum conservation~\cite{BR59,LP73,DPK92}. 
The flux for the attracted species, usually collisional ions, is much harder to obtain. Unlike the 
electron flux, the ion flux depends not only on the field of the macroscopic body but also on 
scattering processes due to the surrounding plasma, which throughout we assume to be quiescent. 
For weak ion collisionalities the charge-exchange enhanced ion flux model proposed by Lampe and 
coworkers~\cite{LGG01,LGS03,SLR04} is usually used. Its validity has been however questioned by 
Tskhakaya and coworkers~\cite{TTS01,TSS01}. We come back to Lampe and coworkers approach below when 
we discuss representative results for our surface model.

Hence, irrespective of the approximations made for the plasma fluxes, the standard approach of 
calculating surface charges is based on three assumptions about the surface physics: 
\begin{center}
\begin{tabbing}
$\bullet$ \= Both ions and electrons reach the surface, even on the\\
          \> microscopic scale.\\
$\bullet$ \> $s_e=s_i=1$ or at least $s_e=s_i$.\\ 
$\bullet$ \> $\tau_e^{-1}=0$ or at least $\tau_e^{-1}\sigma_e\ll s_ij_i^{\rm plasma}=\alpha_R\sigma_e\sigma_i.$ 
\end{tabbing}
\end{center}
We basically challenge all three assumptions.
\begin{figure}[t]
\centering
\includegraphics[width=0.96\linewidth]{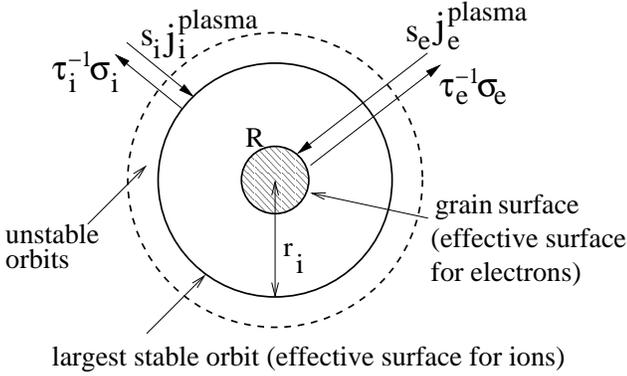}
\caption{Illustration of the surface model for the charging of
a dust particle with radius $R$ in a gas discharge. At
quasi-stationarity, surface charges $\sigma_{e,i}$ bound at
$r_e\approx R$ and $r_i \gtrsim r_e$,
respectively, balance the collection flux $s_{e,i}j^{\rm plasma}_{e,i}$
with the respective desorption flux $\tau_{e,i}^{-1}\sigma_{e,i}$, where
$s_{e,i}$ and $\tau_{e,i}$ denote, respectively, sticking
coefficients and desorption times~\cite{BFK08}.}
\label{SurfaceModel}
\end{figure}

First, electrons and ions should be bound in surface states. Because of differences
in the potential energy, mass, and size the spatial extension of the electron and 
ion bound states, and thus the average distance of electrons and ions from the 
boundary, is expected to be different. On the microscopic scale, electrons and ions 
trapped to the surface should be spatially separated. 

Second, $s_e=s_i$ is quite unlikely. Usually, 
heavy particles, such as ions, couple rather strongly to vibrational excitations of the 
boundary~\cite{KG86,BR92}. They can thus dissipate energy very efficiently which usually 
leads to a large sticking coefficient. Light particles, like electrons, on the other 
hand, couple only very weakly to vibrations of the solid. On this basis, we would 
expect $s_e\ll s_i$. To what extend the coupling to other elementary excitations of the 
boundary (plasmons, electron-hole pairs, ...) can compensate for the inefficient coupling 
to lattice vibrations is part of our investigations. 

Third, if ions and electrons are indeed spatially separated, the two rate equations should 
be in fact interpreted as flux balances on two different effective surfaces (viz: the two 
closed circles in Fig.~\ref{SurfaceModel}). In that case, 
$\alpha_R\sigma_i\sigma_e\ll \sigma_{e,i}/\tau_{e,i}$ and the surface charge $Z_p$ would
be determined by balancing on the grain surface the electron desorption flux,
$\tau_e^{-1}\sigma_e$, with the electron collection flux, $s_ej_e^{\rm plasma}$. The 
corresponding balance of ion fluxes, to be taken on an effective surface surrounding 
the grain, would then yield a partial screening charge $Z_i$. Within this scenario, we 
would thus obtain 
\begin{eqnarray}
Z_p&=&4\pi r_e^2\cdot (s\tau)_e\cdot j_e^{\rm plasma}(Z_p)~,
\label{OurZp}\\
Z_i&=&4\pi r_i^2\cdot (s\tau)_i\cdot j_i^{\rm plasma}~,
\label{OurZi}
\label{Charge}
\end{eqnarray}
with $r_e\approx R$ and $r_i\gtrsim r_e$. 

The surface physics is now encoded in $(s\tau)_{e,i}$. These products depend on the material 
and the plasma. They could be used as adjustable parameters. A justification of the 
assumptions, however, made in deriving Eqs.~(\ref{OurZp}) and~(\ref{OurZi}) can only come 
from a microscopic calculation of $(s\tau)_{e,i}$. 

For electrons, various aspects of this 
calculation will be discussed in the following sections. 

\subsection{Semi-microscopic approach}

Before we discuss the complete microscopic calculation of $s_e$ and $\tau_e$ we 
summarize the semi-microscopic approach taken in Ref.~\cite{BFK08}. This prepares
the grounds for a microscopic thinking and demonstrates that Eqs.~(\ref{OurZp})
and~(\ref{OurZi}) give results which compare favorable with experimental data.

The approach we adopted in Ref.~\cite{BFK08} is based on a quantum mechanical 
investigation of the bound states of a negatively charged particle in a gas discharge. 
For that purpose, we considered the classical interaction between an electron (ion) 
with charge $-e$ ($+e$) and a spherical particle with radius $R$, dielectric constant 
$\epsilon$, and charge $Z_p$. The interaction potential contains then a short-ranged 
polarization-induced part arising from the electric boundary conditions at the grain 
surface -- the classical image potential -- and a long-ranged Coulomb tail due to the 
particle's charge~\cite{Boettcher52,DS87}. 

The polarization-induced part of the potential will be discussed from a quantum-mechanical
point of view in appendix A. Concerning the Coulomb tail we may add that it arises from the 
interaction between the approaching electron and the electrons already residing on the 
grain. From many-body theory it is known that this interaction can be rather involved 
because the attached electrons may respond dynamically~\cite{VR93}. We neglect this 
possibility. The Coulomb part is then simply the potential of a sphere (plane) with 
charge $Z_p$. This is equivalent to a meanfield approximation for the electron-electron 
interaction.

Measuring distances from the grain surface in units of $R$ and energies in units
of $\bar{U}$, the interaction energy at $x=r/R-1>x_b$, where $x_b$ is a lower 
cut-off, below which the grain boundary cannot be described as a perfect surface 
anymore, reads
\begin{eqnarray}
V_{e,i}(x)&=&\pm\frac{1}{1+x}-\frac{\xi}{x(1+x)^2(2+x)}
\nonumber\\
&\approx&\left\{\begin{array}{ll}
1-\xi/2x & \mbox{electron}\\
-1/(1+x) & \mbox{ion}
\end{array}\right.
\label{V(x)}
\end{eqnarray}
with $\xi=(\epsilon-1)/2(\epsilon+1)Z_p$.

\begin{figure}[t]
\vspace{5mm}
\includegraphics[width=0.96\linewidth]{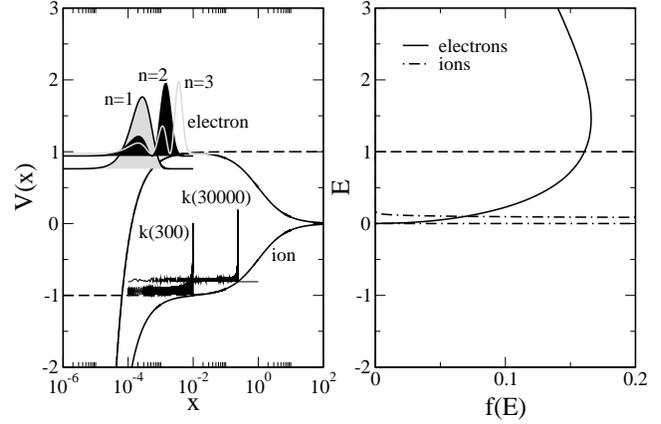}
\caption{Left panel: Potential energy for an electron (ion) in the
field of a MF particle ($R=1~\mu m$, $Z=1500$)~\cite{KRZ05}
and representative probability distributions, $|u(x)|^2$,
shifted to the binding energy and maxima normalized to one.
Dashed lines denote the potentials used in the
Schr\"odinger equations. Note, the finite ion radius $r_i^{size} \sim \AA$
forces the ion wavefunctions to vanish at $x\approx 10^{-4}$.
Right panel: Bulk energy distribution functions for the
$100Pa$ neon discharge hosting the particle~\cite{KRZ05}:
$k_BT_e=6.3eV$, $k_BT_i=0.026eV$, and $n_e=n_i=0.39\times 10^{9}cm^{-3}$.
}
\label{KhrapakPot}
\end{figure}

The second line in Eq.~(\ref{V(x)}) is an approximation which describes
the relevant parts of the potential very well and permits an analytical
calculation of the surface states. In Fig.~\ref{KhrapakPot} we plot
$V_{e,i}(x)$ for a melamine-formaldehyde (MF) particle ($\epsilon=8$,
$R=1~\mu m$, and $Z_p=1500$) embedded in a $100Pa$ neon discharge with
plasma density $n_e=n_i=0.39\times 10^9~cm^{-3}$, ion temperature
$k_BT_i=0.026~eV$, and electron temperature $k_BT_e=6.3~eV$~\cite{KRZ05}.
From the electron energy distribution, $f_{e}(E)$, we see that
the discharge contains enough electrons which can overcome the Coulomb
barrier of the dust particle. These electrons may get bound in the
polarization-induced short-range part of the potential, well described 
by the approximate expression, provided they can get rid of their 
kinetic energy. Ions, on the other hand, being cold
(see $f_i(E)$ in Fig.~\ref{KhrapakPot}) and having a finite
radius $r^{size}_i/R=x_i^{size}\gtrsim 10^{-4}$, cannot explore the
potential at short distances. For them, the long-range Coulomb tail
is most relevant, which is again well described by the approximate
expression.

Writing for the electron eigenvalue $\varepsilon^{e}=1-\alpha_e\xi/4k^2$
with $\alpha_e=(\epsilon-1)R/4(\epsilon+1)a_B$ and for the ion eigenvalue
$\varepsilon^{i}=-\alpha_i/2k^2$ with $\alpha_i=m_iRZ_p/m_ea_B$, where
$m_e$ and $m_i$ are the electron and ion mass,
respectively, the radial Schr\"odinger equations with the approximate potentials
read
\begin{eqnarray}
\frac{d^2u^{e,i}}{dx^2}+\bigg[-\frac{\alpha_{e,i}^2}{k^2}+\tilde{V}_{e,i}(x)
-\frac{l(l+1)}{(1+x)^2}
\bigg]u^{e,i}=0
\label{SE}
\end{eqnarray}
where $\tilde{V}_e(x)=2\alpha_e/x$ and $\tilde{V}_i(x)=2\alpha_i/(1+x)$.

For bound states, the wavefunctions have to vanish for $x\rightarrow\infty$.
The boundary condition at $x_b$ depends on the potential for $x\le x_b$, that
is, on the potential within the solid (which is different for electrons and 
ions). Matching the solutions for $x<x_b$ and $x>x_b$ at $x=x_b$ leads to a 
secular equation for $k$. Ignoring the possibility that electrons and ions may
also enter the solid, we set $\tilde{V}_{e,i}(x\le x_b)=\infty$ with $x_b=0$ 
for electrons and $x_b=x_i^{size}$ for ions. For electrons we thereby restrict 
ourselves to weakly bound polarization-induced surface states, neglecting 
strongly bound crystal-induced surface states which, in general, may also 
occur~\cite{Spanjaard96}. As explained in the next section, we expect them 
to be of minor importance for physisorption of electrons. 

The electron Schr\"odinger equation with the hard boundary condition at $z=0$
is equivalent to the radial Schr\"odinger equation for the hydrogen atom. Hence $k$
is an integer $n$. Because (for bound electrons) $x\ll 1$ and $\alpha_e\gg 1$, 
the centrifugal term is negligible. We consider therefore only states with
$l=0$. The eigenvalues are then $\varepsilon^e_n=1-\alpha_e\xi/4n^2$ and the 
wavefunctions read 
\begin{eqnarray}
u_{n,0}^e(x)&\sim&v_{n,0}(\bar{z})
\nonumber\\
&=&\bar{z}\exp(-\bar{z}/2)(-)^{n-1}(n-1)!L^{(1)}_{n-1}(\bar{z})
\label{vn0}
\end{eqnarray}
with $\bar{z}=2\alpha_ex/n$ and $L^{(1)}_n(\bar{z})$ associated Laguerre
polynomials.

The probability densities $|u_{n,0}^e(x)|^2$ for the first
three states are plotted in Fig.~\ref{KhrapakPot}. As can be seen, electron
surface states are only a few $\AA$ngstroms away from the grain boundary. At
these distances, the spatial variation of $V_e(x)$ is comparable to the
de-Broglie wavelength of electrons approaching the particle.
More specifically, for $k_BT_e=6.3~eV$, $\lambda_e^{dB}/R\approx |V_e/V_e'|
\approx 10^{-4}$. Hence, the trapping of electrons at the surface of the particle 
has to be described quantum-mechanically.

The solutions of the ion Schr\"odinger equation are Whittaker functions,
$u^i_{k,l}(x)=W_{k,l+1/2}(\bar{x})$ with $\bar{x}=2\alpha_i(1+x)/k$ and
$k$ determined from $u^i_{k,l}(x_i^{size})=0$. However, since
$k\gg 1$ and $\bar{x}\gg 1$, it is
very hard to work directly with $W_{k,l+1/2}(\bar{x})$.
It is easier to use the method of comparison equations~\cite{Richardson73}
and to construct uniform approximations for $u^i_{k,l}(x)$
with the radial Schr\"odinger equation for the hydrogen atom as a
comparison equation. The method can be applied for any $l$. Here
we give only the result for $l=0$:
\begin{eqnarray}
u_{k,0}^i(x)\sim v_{n,0}(\bar{z})/\sqrt{dz/dx}
\end{eqnarray}
with $v_{n,0}(\bar{z})$
defined in Eq.~(\ref{vn0}) and $\bar{z}=2\alpha_i z(x)/n$. The mappings $z(x)$
and $k(n)$ can be constructed from the phase-integrals of the
two Schr\"odinger equations. 

In Fig.~\ref{KhrapakPot} we show $|u_{k,0}^i(x)|^2$ for $k(300)$ and
$k(30000)$. Note, even the $k(30000)$ state is basically at the bottom 
of the potential. This is a consequence of $\alpha_i\gg 1$ which leads 
to a continuum of states below the ion ionization threshold at $\varepsilon=0$.
We also note that $|u_{k(n),0}^i(x)|^2$ peaks for $n\gg 1$ just below
the turning point. Hence, except for the lowest states, which we expect
to be of little importance, ions are essentially trapped in classical 
orbits deep in the sheath of the grain. This will be also the
case for $l>0$. That ions behave classically is not unexpected because
for $k_BT_i=0.026~eV$ their de-Broglie wavelength is much smaller then
the scale on which the potential varies for $x>10^{-3}$:
$\lambda_i^{dB}/R\approx 10^{-5}\ll |V_i/V_i'|\approx 1$. Thus, the 
interaction between ions and the particle is classical. 

Nevertheless it can be advantageous to describe ions quantum-mechanically and to use the
method of comparison equations, which is an asymptotic technique, to perform the 
calculation in the semiclassical regime. Since the ion dynamics and kinetics is 
beyond the scope of this paper, we do not give more mathematical details about the
solution of the ion Schr\"odinger equation. We mention however that many years ago 
Liu~\cite{Liu69} pursued a quantum-mechanical description of the collisionless ion 
dynamics around electric probes. But he found no followers. 

A model for the charge of the grain which takes surface states into account can
now be constructed as follows. Within the sheath of the particle, the density of 
free electrons (ions) is much smaller than the density of bound electrons (ions).
In that region, the quasi-stationary charge (again in units of $-e$) is thus approximately
given by
\begin{eqnarray}
Z(x)=4\pi R^3\!\int_{x_b}^x\!\!dx' \big(1+x'\big)^2 \bigg[n^b_e(x')-n^b_i(x')
\bigg]
\label{Zintegral}
\end{eqnarray}
with $x<\lambda^D_i=\sqrt{k_BT_i/4\pi e^2 n_i}$, the ion Debye length, which we take as 
an upper cut-off, and $n^b_{e,i}$ the density of bound electrons and ions. For the 
plasma parameters used in Fig.~\ref{KhrapakPot}, $\lambda^D_i\approx 60\mu m$.
The results for the surface states presented above suggest to express the density 
of bound electrons by an electron surface density: 
\begin{eqnarray}
n^b_e(x)\approx\sigma_e\delta(x-x_e)/R
\label{nb}
\end{eqnarray} 
with $x_e\approx x_b\approx 0$ and $\sigma_e$ the quasi-stationary solution of
of Eq.~(\ref{REqe}) without the recombination term. Equation~(\ref{REqe}) is thus still 
interpreted as a rate equation on the grain surface. We will argue below that once the 
grain has collected some negative charge, not necessarily the quasi-stationary one, 
there is a critical ion orbit at $x_i\sim 1-10 \gg x_e$ which prevents ions from hitting 
the particle surface. Thus, the particle charge obtained from Eq.~(\ref{Zintegral}) is 
simply $Z_p\equiv Z(x_e<x<x_i)$. Inserting Eq. (\ref{nb}) into Eq. (\ref{Zintegral}) and 
integrating up to $x$ with $x_e<x<x_i$ leads to Eq. (\ref{OurZp}), the expression
for the particle charge deduced from the rate equations (\ref{REqe}) and (\ref{REqi}) 
under the assumption that ions do not reach the grain surface on the microscopic scale.

For an electron to get stuck at (to desorb from) a surface it has to loose (gain) 
energy at (from) the surface~\cite{KG86}. This can only occur through inelastic 
scattering with the grain surface. To calculate the product $(s\tau)_e$ requires 
therefore a microscopic description of energy relaxation at the grain surface. 
This will be discussed in the next section. In Ref.~\cite{BFK08} we invoked the 
phenomenology of reaction rate theory and approximated $(s\tau)_e$ by 
\begin{eqnarray}
(s\tau)_e=
\frac{h}{k_BT_s}\exp\bigg[\frac{E_e^d}{k_BT_s}\bigg]~,
\label{stau}
\end{eqnarray}
where $h$ is Planck's constant, $T_s$ is the surface temperature, and $E_e^d$ is the 
electron desorption energy, that is, the binding energy of the surface state from
which desorption most likely occurs~\cite{KG86}. The great virtue of this 
equation is that it relates a combination of kinetic coefficients, which depend on the
details of the inelastic (dynamic) interaction, to an energy, which can be deduced from
the static interaction alone. Kinetic considerations are thus reduced to a minimum.
They are only required to identify the relevant temperature and the state from which 
desorption most probably occurs. In the next section we will show, for a particular 
model, how Eq.~(\ref{stau}) can be obtained from a microscopic theory. Its range of 
validity will then become also clear.

Equation~(\ref{OurZp}) is a self-consistency equation for $Z_p$. Combined with 
Eq.~(\ref{stau}), and approximating the electron flux $j_e^{\rm plasma}$ from 
the plasma by the orbital motion limited flux, 
\begin{eqnarray}
j_e^{\rm OML}=n_e\sqrt{k_BT_e/2\pi m_e}\exp[-Z_pe^2/Rk_BT_e]~,
\label{jeOML}
\end{eqnarray}
which is reasonable, because, on the plasma scale, electrons are repelled from the
grain surface, the grain charge is given by 
\begin{eqnarray}
Z_p=4\pi R^2\frac{h}{k_BT_s}e^{E_e^d/k_BT_s}j_e^{\rm OML}(Z_p)~.
\label{Zpfinal}
\end{eqnarray}
Thus, in addition to the plasma parameters $n_e$ and $T_e$, the charge depends on 
the surface parameters $T_s$ and $E_e^d$.
\begin{figure}[t]
\vspace{5mm}
\includegraphics[width=0.96\linewidth]{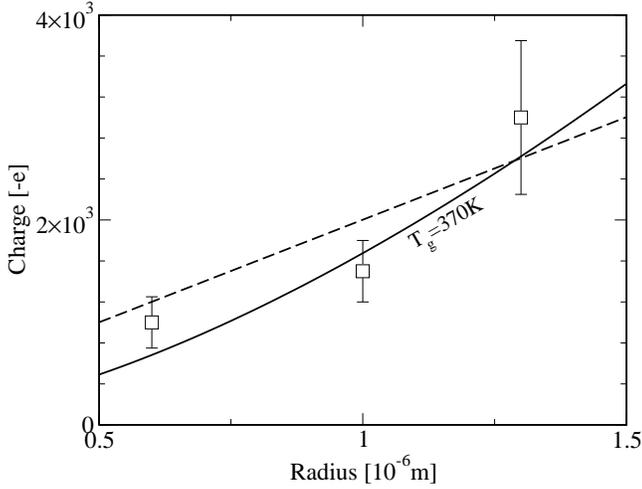}
\caption{Radius dependence of the charge of a MF particle in
the bulk of a neon discharge at $p=100Pa$~\cite{KRZ05}. The
plasma parameters are the same as in Fig.~\ref{KhrapakPot}. The
solid line denotes the charges deduced from Eq.~(\ref{Zpfinal})
and the dashed line gives the charges obtained from
$j_e^{\rm OML}=j_i^{\rm OML}+j_i^{\rm CX}$ with
$\sigma_{\rm cx}=10^{-14}cm^{-2}$.
}
\label{KhrapakData}
\end{figure}

Without a microscopic theory for the inelastic electron-grain interaction, a plausible
estimate for $E_e^d$ has to be found from physical considerations alone. Since by necessity
the electron comes very close to the grain surface (see Fig.~\ref{KhrapakPot}) it 
will strongly couple to elementary excitations of the grain. Depending on the material
these may be bulk or surface phonons, bulk or surface plasmons, or internal electron-hole 
pairs. For any realistic description of the potential for $x\le x_b$ the electron 
wavefunction leaks into the solid, the electron will therefore quickly relax to the lowest 
surface bound state. The microscopic model for electron energy relaxation at metallic 
boundaries presented in the next section turns out to even work for an infinitely 
high barrier. Taking the $n=1$ state for $\tilde{V}_{e,i}(x\le x_b)=\infty$ as an 
approximation to the lowest surface bound state, it is reasonable to expect
\begin{eqnarray}
E_e^d\approx(1-\varepsilon^e_1)\bar{U}=\frac{R_0}{16}\bigg(\frac{\epsilon-1}{\epsilon+1}\bigg)^2~,
\end{eqnarray}
which, for an MF particle with $\epsilon=8$, leads to $E_e^d\approx 0.5eV$. The particle temperature
cannot be determined in a simple way. It depends on the balance of heating and cooling fluxes
to-and-fro the particle and thus on additional surface parameters~\cite{SKD00}. We use $T_s$ therefore
as an adjustable parameter. To reproduce, for instance, with Eq.~(\ref{Zpfinal}) the charge of the
particle in Fig.~\ref{KhrapakPot}, $T_s=370~K$ implying $(s\tau)_e\approx 10^{-6}~s$.

In Fig.~\ref{KhrapakData} we plot the radius dependence of the charge of a MF particle in 
the $100Pa$ neon discharge specified in the caption of Fig.~\ref{KhrapakPot}. More results
are given in~\cite{BFK08}. Since the plasma parameters are known the only adjustable parameter 
is the surface temperature. Using $T_s=370K$ we find excellent agreement between theory and 
experiment. For comparison we also show the charges obtained from Eq.~(\ref{OrdinaryApp}), 
approximating the ion plasma flux by
\begin{eqnarray}
j_i^{\rm plasma}=j_i^{\rm OML}+j_i^{\rm cx}~,
\end{eqnarray}
where 
\begin{eqnarray}
j_i^{\rm OML}=n_i\sqrt{k_BT_i/2\pi m_i}[1+Z_pe^2/R k_BT_i]
\label{jiOML}
\end{eqnarray}
is the orbital motion limited ion flux and~\cite{KRZ05} 
\begin{eqnarray}
j_i^{\rm cx}=n_i(0.1\lambda^D_i/l_{\rm cx})
\sqrt{k_BT_i/2\pi m_i}(Z_pe^2/R k_BT_i)^2
\label{jiCX}
\end{eqnarray}
is the ion flux originating from the release of trapped ions due to charge-exchange 
scattering as suggested by Lampe and coworkers~\cite{LGG01,LGS03,SLR04}. The 
scattering length $l_{\rm cx}=(\sigma_{\rm cx} n_g)^{-1}$ with 
$\sigma_{\rm cx}=10^{-14}cm^2$ the scattering cross section and $n_g=p/k_BT_g$ the 
gas density. Clearly, the radius dependence of the grain charge seems
to be closer to the nonlinear dependence obtained from Eq.~(\ref{Zpfinal}) than
to the linear dependence resulting from 
\begin{eqnarray}
j_e^{\rm OML}=j_i^{\rm OML}+j_i^{\rm cx}~, 
\end{eqnarray}
indicating that the surface model we propose captures at least some of the physics 
correctly which is responsible for the formation of surface charges. 

In order to derive Eq.~(\ref{Zpfinal}) from Eq.~(\ref{Zintegral}) we had to assume that 
once the particle is negatively charged ions are trapped far away from the grain surface.
Treating trapping of ions in the field of the grain as a physisorption process suggests 
this assumption, which is perhaps counter-intuitive. Similar to an electron, an ion gets 
bound to the grain only when it looses energy. Because of its low energy and the long-range 
attractive ion-grain interaction, the ion will be initially bound very close to the 
ion ionization threshold (see Fig.~\ref{KhrapakPot}). The coupling to the elementary 
excitations of the grain is thus negligible and only inelastic processes due to the 
plasma are able to push ions to lower bound states. Since the interaction is classical,
inelastic collisions, for instance, charge-exchange scattering between ions and atoms, 
act like a random force. Ion energy relaxation can be thus envisaged as a de-stabilization
of orbits. This is in accordance to what Lampe and coworkers assume~\cite{LGG01,LGS03,SLR04}. 
In contrast to them, however, we~\cite{BFK08} expect orbits whose spatial extension is 
smaller than the scattering length to be stable because the collision probability during one 
revolution becomes vanishingly small. For a circular orbit, a rough estimate for the critical 
radius is 
\begin{eqnarray}
r_i=R(1+x_i)=(2\pi\sigma_{\rm cx} n_g)^{-1}
\end{eqnarray}
which leads to $x_i\sim 5.7\gg x_e \sim 0$ when we use the parameters of the neon discharge of 
Fig.~\ref{KhrapakPot} and $\sigma_{\rm cx}=10^{-14}~cm^2$. 

Although the approach of Lampe {\it et al.}~\cite{LGG01,LGS03,SLR04} shows a pile-up of 
trapped ions in a shell of a few $\mu m$ radius enclosing the grain, they would not expect 
a relaxation bottleneck. This point can be only clarified with a detailed investigation 
of the ion dynamics and kinetics in the vicinity of the grain, including electron-ion
recombination. As mentioned before, despite 
the classical character of the ion dynamics, a quantum-mechanical treatment, similar 
to the one we will present in the following sections for electrons, is possible
and perhaps even advantageous because it treats closed (bound surface states)
and open ion orbits (extended surface states) on the same footing. In addition, 
energy barriers due to the angular motion are easier to handle 
in a quantum-mechanical context. In fact, Lampe and coworkers neglect these energy
barriers whereas Tskhakaya and coworkers~\cite{TTS01,TSS01} believe that this 
approximation overestimates $j_i^{\rm cx}$. In reality, they claim, $j_i^{\rm cx}$ 
is much smaller. If this is indeed the case, the condition $j_e^{\rm OML}=j_i^{\rm OML}+j_i^{\rm cx}$ 
would yield charges which are much closer to the orbital-motion limited ones 
and thus far away from the experimentally measured charges. 

Pushing the assumption of a critical ion orbit even further, we assumed in~\cite{BFK08}
that all trapped ions can be subsumed into a single effective orbit as shown in 
Fig.~\ref{SurfaceModel}. We then obtained an intuitive expression for the number of ions 
accumulating in the vicinity of the grain, that is, for its partial screening charge. 
For that purpose we modelled the ion density $n_i^b$ accumulating in the vicinity of the 
critical orbit by a surface density $\sigma_i$ which balances at $x_i$ the ion collection 
flux $s_ij_i^{\rm plasma}$ with the ion desorption flux $\tau_i^{-1}\sigma_i$.
Mathematically, this gives rise to a rate equation similar to~(\ref{REqi}), 
with the recombination term neglected and interpreted as a rate equation at 
$r=r_i$. Although Eq.~(\ref{stau}) assumes excitations of the grain to be responsible 
for sticking and desorption we expect a similar expression (with $E_e^d$,
$T_s$ replaced by $E_i^d$, $T_g$) to control the density of trapped ions. Integrating
~(\ref{Zintegral}) up to $x$ with $x_i<x<\lambda^D_i$ we then obtain 
$Z(x_i<x<\lambda^D_i)=Z_p-Z_i$ with
\begin{eqnarray}
Z_i=4\pi R^2(1+x_i)^2
\frac{h}{k_BT_g}e^{E_i^d(Z_p)/k_BT_g}j^B_i~
\label{Zion}
\end{eqnarray}
the number of trapped ions. Since the critical orbit is near the sheath-plasma boundary, 
it is fed by the Bohm ion flux 
\begin{eqnarray}
j^B_i=0.6n_i\sqrt{k_BT_e/m_i}~. 
\end{eqnarray}
The ion desorption energy is the negative of the binding energy of the critical orbit,
\begin{eqnarray}
E_i^d(Z_p)=-V_i(x_i)\bar{U}(Z_p)=4\pi\sigma_{\rm cx}a_B n_g Z_pR_0~,
\end{eqnarray}
and depends strongly on $Z_p$ and $x_i$. For the situation shown in Fig.~\ref{KhrapakPot}, 
we obtain $E_i^d\approx 0.39eV$ and $(s\tau)_i\approx 10^{-8}~s$ when we 
use $T_g=T_s=370~K$, the particle temperature which reproduces $Z_p\approx 1500$. The 
ion screening charge is then $Z_{i}\approx 12\ll Z_p$ which is the order of magnitude
expected from molecular dynamics simulations~\cite{CK94}. Thus, even when the particle 
charge is defined by $Z(x_i<x<\lambda^D_i)$ it is basically given by $Z_p$.

From the surface model we would expect $(s\tau)_e\sim 10^{-6}s$ to produce particle
charges $Z_p$ of the correct order of magnitude. Since the particle temperature $T_s$ 
is unknown, it can be used as an adjustable parameter. The calculated $Z_p$ can 
thus be always made to coincide with the measured charge. The particle temperature 
has to be of course within physically meaningful bonds. Recently, the particle 
temperature (but unfortunately not the particle charge) has been measured~\cite{MBK08}. 
There is thus some hope that in the near future $Z_p$ and $T_s$ will be simultaneously 
measured. Finally, let us point out that, because ions are in our model bound a few
microns away from the surface, we obtain $(s\tau)_i<(s\tau)_e$, in agreement with
the phenomenological fit performed in~\cite{KDK04}. 

\section{Physisorption of electrons}

In the previous section we described a microscopic, physi- sorption-inspired model
for the charging of a dust particle in a plasma which avoids the unrealistic 
treatment of the grain as a perfect absorber. Within this model the charge and 
partial screening of a dust particle can be calculated without relying on the 
condition that the total electron plasma flux balances on the grain surface the
total ion plasma flux. Instead, two flux balance conditions are individually 
enforced on the two effective surfaces shown in Fig.~\ref{SurfaceModel} (solid circles). 
The quasi-stationary particle charge $Z_p$ is then given by the number of 
electrons \textquotedblleft quasi-bound\textquotedblright~in the polarization
potential of the grain and the screening charge $Z_i$ is approximately given by the
number of ions \textquotedblleft quasi-trapped\textquotedblright~in the largest
stable closed ion orbit (which defines an effective surface for ions and subsumes,
within our model, all trapped ions into a single effective orbit). 

The physisorption kinetics at the grain boundary, that is, the sticking in and the
desorption from (external) surface states due to inelastic scattering processes, is
encoded in the products $(s\tau)_{e,i}$ which we approximated by phenomenological 
expressions of the form~(\ref{stau}). For electrons, we now take a closer look at 
what happens on the surface microscopically. First, we will discuss the microphysics
qualitatively. Then we will perform an exploratory quantum-mechanical calculation of 
$s_e$ and $\tau_e$ using a simple one-dimensional model for the electronic properties
of the surface which allows us to do large portions of the calculation analytically. 
Finally, we will critically assess the results of the calculation turning thereby its 
shortcomings into a list of to-do's.

In principle, trapping and de-trapping of ions in the surface-induced Coulomb potential of 
the grain can be also understood as a physisorption process. However, the quantum-mechanical 
approach we will use for electrons has then to be pushed to the semi-classical regime 
appropriate for ions. In addition, not surface- but plasma-based inelastic scattering 
processes will turn out to control ion energy relaxation. Although conceptually very close, 
mathematically the calculation of $s_i$ and $\tau_i$ is quite different from the 
calculation of $s_e$ and $\tau_e$. It is therefore beyond the scope of this paper. In the
concluding section we may however add a few remarks about ions.

\subsection{Qualitative considerations}
\label{SubSec:11}

The surface of a $\mu m$-sized grain is large enough to contain sizeable spatial 
regions (facets) isomorphous to crystallographic planes. Except specific features  
arising from the finite extend of the facets, whose influence diminishes 
with the facet size, the electronic properties of the facets 
%can be expected to 
resemble the electronic properties of (infinitely extended) planar surfaces. 
In particular, like ordinary surfaces, facets should support surface states to which 
electrons approaching the grain from the plasma may get bound and then re-emitted when 
they dynamically interact with the elementary excitations of the grain.
\begin{figure}[t]
\centering
\includegraphics[width=0.96\linewidth]{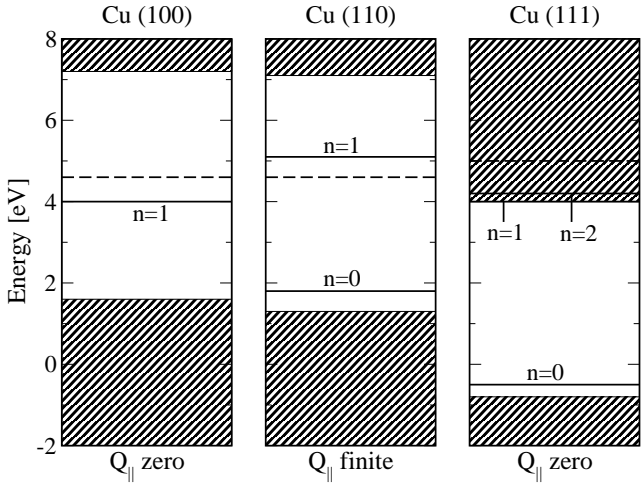}
\caption{Schematic electronic structure after~\cite{JDK86} for a copper (100), a copper
(110), and a copper (111) surface, respectively, for the lateral momentum where the projected
energy gap is largest: $|\vec{Q}|=0$ for Cu (100) and Cu (111); $|\vec{Q}|=1.2\AA^{-1}$ in
$[\bar{1}10]$ direction for Cu (110). Shaded areas denote the projected bulk band structure
and thick solid lines indicate crystal-induced ($n=0$) and polarization-induced
($n\ge 1$) surface states. The zero of the energy axis is the Fermi energy
and the vacuum level is given by the dashed line.}
\label{Copper}
\end{figure}

Each facet may give rise to two types of surface states: (i) Crystal-induced surface states
due to the abrupt appearance (from the plasma electron's point of view) of a periodic 
potential inside the grain and (ii) polarization-induced image states, on which the considerations
of the previous section were based. Compared to the binding energy of image states, the 
binding energy of crystal-induced surface states is very large. Instead of a few tenth 
of an electron volt, it is typically a few electron volts. As a result, the center of 
gravity of crystal-induced surface states is much closer to the surface than the center 
of gravity of image states.

Based on the experimental results of~\cite{JDK86} we show in Fig.~\ref{Copper} as an example
the schematic electronic structure of three copper surfaces, respectively, for the lateral 
momentum where the projected energy gap is largest. The electronic structure for a given 
orientation changes with momentum (not shown) but for all orientations, and that is the 
point we want to make, surface states exist\footnote{More precisely, surface states exist
in some parts of the surface Brillouin zone~\cite{JDK86}.}, in addition to projected bulk 
states, and may thus participate in a physisorption process. For dielectric surfaces the 
electronic structure is quite similar although the details and physical origin of the 
states is different~\cite{Spanjaard96}. 

%Image states are energetically far away from the Fermi energy of the material and thus 
%usually unoccupied. They may be therefore more important for physisorption than 
%crystal-induced surface states which, being close to the Fermi energy of the material, 
%are very often already occupied by crystal electrons.

\begin{figure}[t]
\centering
\includegraphics[width=0.90\linewidth]{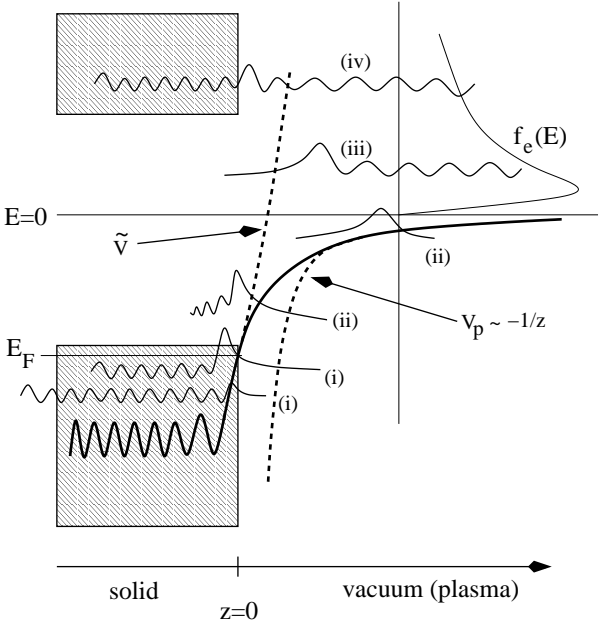}
\caption{Schematic drawing of the potential energy at a surface (plasma boundary) such as
copper (100) or copper (110) and representative wavefunctions for volume states (i), bound
surface states (ii), unbound surface states (iii), and free states (iv). Shaded areas denote
projected bulk states, $E_F$ is the Fermi energy of the solid, $E=0$ is the vacuum (i.e. floating)
level, and $f_e(E)$ is the plasma electron's Boltzmann distribution function. The dashed lines indicate
the approximate potentials defining the simplified planar model of subsection~\ref{SectionModel}
on which the calculation of $s_e$ and $\tau_e$ is based we describe, respectively, in
subsection~\ref{SectionStickCoeff} and~\ref{SectionDesorbTime}.
}
\label{SurfaceStates}
\end{figure}

An ab-initio modeling of surface states is complex and computationally
expensive, even for planar surfaces (see for instance~\cite{HFH98}). Fortunately, the
essential physics can be understood within simple one dimensional models which assume
the potential energy to vary only normally to the surface ($z-$direction) as illustrated
in Fig.~\ref{SurfaceStates}. Inside the material ($z<0$) the potential has the periodicity 
of the crystal. It may thus lead to an energy gap on the surface. Outside the material, 
the potential gives rise to a barrier which merges at large distances with the asymptotics 
of the image potential $V_p(z)\sim -1/z$. Its physical origin are exchange and correlation 
effects which, on the one hand, contribute to the confinement of electrons inside the material
and, on the other hand, cause the attraction of external electrons to the surface. A simple 
microscopic model for the image potential~\cite{RM72,EM73} is given in appendix A.

The situation shown in Fig.~\ref{SurfaceStates} is the most favorable one for physisorption 
of electrons. The vacuum (plasma) potential, which is the zero of the energy scale, is in the middle 
of a large energy gap. Four main classes of states can then be distinguished: (i) Volume 
states periodic inside the material and exponentially decaying into the vacuum (plasma). They 
exist for energies where bulk states are also allowed. Close to band edges they may have an 
increased weight near the surface in which case they are surface resonances. (ii) Bound
surface states, that is, states decaying exponentially into the material 
and the vacuum. They appear in regions of negative energies where bulk states are absent:
Weakly bound image states close to the vacuum potential and strongly bound crystal-induced 
surface states close to the Fermi energy. Crystal-induced surface states may have tails on
the material side strongly oscillating with the crystal periodicity, while the tails of 
image states may only weakly respond to the crystal potential. (iii)
Unbound surface states for positive energies inside the gap. They are free on the vacuum and 
bound on the material side. The periodic crystal potential may also not affect these states 
very much. (iv) States which are free on both sides. Inside the material they oscillate with 
the lattice periodicity while outside the material their oscillations have to fit the
surface potential. In the vicinity of the surface this class of states may also have a peak.

Of particular importance for sticking and desorption are transitions between bound and
unbound surface states due to inelastic scattering with elementary excitations of the 
boundary. The elementary excitations can be phonons, plasmons, and electron-hole pairs. 
The latter two cases are excitations involving volume states. 

The potential plotted in Fig.~\ref{SurfaceStates} is for an uncharged surface. An 
electron approaching a plasma boundary is of course also subject to the Coulomb repulsion 
due to the electrons already residing on the surface. In the meanfield approximation, 
however, this repulsion leads only to a barrier whose height is the floating energy $\bar{U}$. 
Only an electron with an energy larger than $\bar{U}$ has a chance to come close enough to 
the surface to feel the attractive part of the potential. For an electron
bound in this part of the potential, on the other hand, the Coulomb barrier merely sets 
the ionization threshold. Thus, as long as the Coulomb repulsion is treated in meanfield 
approximation, the Coulomb term drops out from the considerations provided we shift the 
zero of the energy axis to $\bar{U}$, that is, by simply measuring energies with respect to
the floating energy (Coulomb barrier) of the surface. If $\bar{U}$ falls inside an energy gap
of the boundary the situation is similar to the one depicted in Fig.~\ref{SurfaceStates}.

\subsection{Simplified planar microscopic model}
\label{SectionModel}

Ideally, a microscopic calculation of $s_e$ and $\tau_e$ for a spherical grain would be based 
on a three-dimensional first-principle electronic structure of the grain surface. 

In view of the discussion of the previous subsection an estimate for the grain's $s_e$ and 
$\tau_e$ may be however also obtained by the following strategy which is most probably simpler 
because it allows to incorporate existing (one-dimensional) empirical pseudo-potentials for 
planar surfaces: (i) Identify the facets on the grain surface and neglect, in a 
first approximation, the finite lateral extension of the facets, that is, work with plane 
waves or Bloch functions in the lateral dimensions. (ii) Use empirical one-dimensional 
potentials for planar surfaces~\cite{CSE99} to calculate for each facet separately bound and 
unbound surface states. (iii) Identify the channels for electron energy relaxation and set up, 
again for each facet separately, a quantum-kinetic scheme for the calculation of $s_e$ and 
$\tau_e$. (iv) Use an appropriate macroscopic spatial averaging scheme to obtain an estimate 
for the grain's $s_e$ and $\tau_e$. 

Despite its approximate nature this strategy is still demanding. To work it out
for a realistic grain is surely beyond the scope of this colloquium. In the exploratory 
calculation of $s_e$ and $\tau_e$ presented below we focused therefore on a single, 
infinitely extended facet, that is, on a planar surface, whose electronic structure we moreover 
did not deduce from an empirical pseudo-potential but from a model potential which is amenable
to analytical treatment while at the same time it retains the essential physics.

Quite generally, the probability with which an electron approaching from the plasma halfspace
$z>0$ the plasma boundary at $z=0$ ends up in a bound surface state (sticking), or with which 
an electron bound to the surface ends up in a free state (desorption) can be obtained from a 
Hamiltonian, 
\begin{eqnarray}
H=H_e+H_s+H_{es}~,
\label{PhysiH}
\end{eqnarray}
where the first term describes the electron motion in the static surface potential,
the second term denotes the free motion of the elementary excitations of the boundary 
controlling electron energy relaxation at the boundary and thus physisorption of electrons, 
and the third term is the coupling between the two.

It is advantageous to express the Hamiltonian~(\ref{PhysiH}) in terms of creation and 
annihilation operators for the (external) electron as well as the (internal) elementary
excitations. For that purpose we use the basis in which $H_e$ is diagonal, that is, the 
eigenstates of the static surface potential $V(z)$.~\footnote{In a realistic calculation 
$V(z)$ should be an empirical pseudo-potential of the type proposed in~\cite{CSE99}.} 
Writing ${\bf r}=({\bf R},z)$ for the electron position, the Schr\"odinger equation 
defining these states reads
\begin{eqnarray}
\bigg(-\frac{\hbar^2}{2m_e}\Delta + V(z)\bigg)\Psi_{\vec{Q}q}(\vec{R},z)=
E_{\vec{Q}q}\Psi_{\vec{Q}q}(\vec{R},z)~.
\label{Schroedinger3D}
\end{eqnarray}
The lateral motion is free and can be separated from the vertical one. Hence, 
\begin{eqnarray}
\Psi_{\vec{Q}q}(\vec{R},z)=\frac{1}{\sqrt{A}}\exp[i\vec{Q}\cdot\vec{R}]\psi_q(z)~,
\label{FreeMotion}
\end{eqnarray}
with $A$ the area of the surface, which is eventually made infinitely large, $\vec{Q}=(Q_x,Q_y)$
a two-dimensional wavevector characterizing the lateral motion of the electron and 
$\psi_q(z)$ the wavefunction for the vertical motion which satisfies the one-dimensional
Schr\"odinger equation (viz: Eq.~(\ref{SE}))
\begin{eqnarray}
\frac{d^2}{dz^2}\psi_q(z)+\frac{2m_e}{\hbar^2}\bigg[E_q-V(z)\bigg]\psi_q(z)=0
\label{SGpsi}
\end{eqnarray}
with $E_q=E_{\vec{Q}q}-\hbar^2Q^2/2m_e$. The quantum number $q$ is an integer $n$ for 
bound and a wavenumber $k$ for unbound surface states. In this basis,
\begin{eqnarray}
H_e=\sum_{\vec{Q}q}E_{\vec{Q}q}C_{\vec{Q}q}^\dagger C_{\vec{Q}q}~,
\label{He}
\end{eqnarray}
where $C_{\vec{Q}q}^\dagger$ creates an electron in the surface state $\Psi_{\vec{Q}q}$ 
with energy 
\begin{eqnarray}
E_{\vec{Q}q}=\hbar^2Q^2/2m_e+E_q~. 
\end{eqnarray}
\begin{table}
\caption{Dielectric constant $\epsilon$, Debye Energy $k_BT_D$, and the energy separations
between the four lowest image states for different materials.}
\label{DebyeEnergy}       % Give a unique label
\begin{tabular}{lllll}
\hline\noalign{\smallskip}
material & $\epsilon$ & $k_BT_D[eV]$ & $\Delta E_{21}[eV]$ & $\Delta E_{43}[eV]$\\
\noalign{\smallskip}\hline\noalign{\smallskip}
Cu       & $\infty$ & 0.03  & 0.64 & 0.12 \\
Si       & 12       & 0.057 & 0.46 & 0.09 \\
graphite & 12       & 0.19  & 0.46 & 0.09 \\
$C_{60}$ & 4.5      & 0.016 & 0.26 & 0.05 \\
\noalign{\smallskip}\hline
\end{tabular}
\end{table}

The second and third term in~(\ref{PhysiH}) depend on the kind of elementary excitations 
responsible for energy relaxation and hence on the material. For dielectric materials, 
such as graphite or silicon, the coupling to vibrational modes is most probably the 
main driving force for physisorption of electrons. In particular, lattice vibrations 
should play an important role. Their energy scale is the Debye energy $k_BT_D$. For most 
dielectrics $k_BT_D$ is smaller than the energy spacing of the lowest surface states. 
For image states typical energy separations are given in table~\ref{DebyeEnergy}. When 
crystal-induced surface states or dangling bonds~\cite{Spanjaard96}) are also included
the situation does not change much, it may be even worse. Multiphonon processes could
thus significantly affect physisorption of electrons at dielectric surfaces making it a 
very interesting problem to study.

For metals, on the other hand, electronic excitations, most notably electron-hole pairs,
provide an efficient channel for electron energy relaxation~\cite{NNS86,WJS92}. They
are not created across a large energy gap, as in dielectrics, where they are therefore 
unimportant, but with respect to the Fermi energy of a partially filled band. In metals 
electron-hole pairs can be excited even at room temperature. Physisorption of electrons
at metallic plasma boundaries, whose temperatures are typically not much higher than room
temperature, is thus most likely controlled by the coupling to electron-hole pairs. 

The Fermi energy of a metal is inside a band. Electron-hole pairs are thus excitations
involving volume states. Ignoring, in a first approximation, exchange between these 
states and (bound and unbound) surface states, which should be small because the states
are spatially separated (see Fig.~\ref{SurfaceStates}), electrons occupying these 
two classes of states can be approximately treated as two separate species: External and 
internal electrons, where the latter are responsible for energy relaxation of the former.

Specifically for a metallic plasma boundary, and we will restrict the calculation of
$s_e$ and $\tau_e$ presented in the next two subsections to this particular case, $H_s$ is 
thus the Hamiltonian of a non-interacting gas of electronic quasi-particles with Fermi energy 
$E_F$. Hence,
\begin{eqnarray}
H_s=
\sum_{\vec{K}k}E_{\vec{K}k}D_{\vec{K}k}^\dagger D_{\vec{K}k}~,
\label{Hs}
\end{eqnarray}
with $D_{\vec{K}k}^\dagger$ creating an internal electron in a quasi-particle state 
\begin{eqnarray}
\Phi_{\vec{K}k}(\vec{R},z)=\frac{1}{\sqrt{A}}\exp[i\vec{K}\cdot\vec{R}]\phi_k(z)
\end{eqnarray}
with energy 
\begin{eqnarray}
E_{\vec{K}k}=\frac{\hbar^2K^2}{2m_e}+\tilde{E}_k~.
\end{eqnarray}
In the above expressions we ignored the periodic crystal potential inside the material. 
It could be taken into account using for the lateral motion Bloch states instead of plane 
waves and effective electron masses, possibly different for the lateral and vertical 
motions, instead of the bare electron mass. 
\begin{table}
\caption{Fermi energy $E_F$, Fermi wavenumber $k_F$, and screening
wavenumber $(k_s)_{\rm bulk}$ for various metals~\cite{AshcroftMermin}.}
\label{ScreenPara}       % Give a unique label
\begin{tabular}{llll}
\hline\noalign{\smallskip}
metal & $E_F [eV]$ & $k_F [\AA^{-1}]$ & $(k_s)_{\rm bulk}/k_F$\\
\noalign{\smallskip}\hline\noalign{\smallskip}
Ag & 5.49  & 1.20 & 1.42 \\
Cu & 7.0   & 1.36 & 1.33 \\
Al & 11.7  & 1.75 & 1.17 \\
\noalign{\smallskip}\hline
\end{tabular}
\end{table}

The function $\phi_k(z)$, describing the vertical motion of an internal electron, obeys
a one-dimensional Schr\"odinger equation: 
\begin{eqnarray}
\frac{d^2}{dz^2}\phi_k(z)+\frac{2m_e}{\hbar^2}\bigg[\tilde{E}_k-\tilde{V}(z)\bigg]\phi_k(z)=0~.
\label{SGphi}
\end{eqnarray}
Strictly speaking, the potential $\tilde{V}(z)=V(z)$. But the spatial parts of the potential 
determining, respectively, surface and volume states are different. Working conceptually with 
two separate potentials gives us the flexibility to independently extend the relevant parts 
of the potential such that the calculation of surface and volume states can be most easily 
performed while the essential physics is kept (see dashed lines in Fig.~\ref{SurfaceStates} 
and below for the particular form of the approximate potentials).

For a metallic boundary, the interaction part $H_{es}$ of the Hamiltonian~(\ref{PhysiH}) 
describes the interaction between internal and external electrons. Anticipating a statically 
screened Coulomb interaction, 
\begin{eqnarray}
H_{es}=\frac{1}{2}\sum_{\vec{Q}q\vec{Q}'q'\vec{K}k\vec{K}'k'}
\!\!\!\!V_{\vec{Q}'q'\vec{K}'k'}^{\vec{Q}q~~\vec{K}k}
C_{\vec{Q}q}^\dagger C_{\vec{Q}'q'}D_{\vec{K}k}^\dagger D_{\vec{K}'k'}
\label{Hes}
\end{eqnarray}
with 
\begin{eqnarray}
V_{\vec{Q}'q'\vec{K}'k'}^{\vec{Q}q~~\vec{K}k}&=&\frac{2\pi e^2}{A^2}
\frac{\delta(\vec{Q}-\vec{Q}'+\vec{K}-\vec{K}')}{\sqrt{k_s^2+(\vec{Q}-\vec{Q}')^2}}
\nonumber\\
&\times& I_{q'k'}^{qk}(\vec{Q}-\vec{Q}')
\end{eqnarray}
and 
\begin{eqnarray}
\nonumber\\
I_{q'k'}^{qk}(\vec{Q})\!=\!\!\int\!\!\!dz dz'\!\psi^*_{q}(z)\phi^*_{k}(z')
e^{-d|z-z'|}
\phi_{k'}(z')\psi_{q}(z)~\!\!,
\label{Ielement}
\end{eqnarray}
where $d=\sqrt{k_s^2+\vec{Q}^2}$ and $k_s=(k_s)_{\rm surface}$ is the
screening wavenumber at the surface. Little is known about this parameter except 
that it should be less then the bulk screening wavenumber $(k_s)_{\rm bulk}$ because the 
electron density in the vicinity of the boundary is certainly smaller than in the bulk.
In~\cite{NNS86} it was for instance argued, based on a comparison of experimentally and
theoretically obtained branching ratios for positron trapping at and transmission through
various metallic films that $(k_s)_{\rm surface}\simeq 0.6(k_s)_{\rm bulk}$. Bulk 
screening wavenumbers for some metals are given in table~\ref{ScreenPara}.
\begin{figure}[t]
\centering
\includegraphics[width=0.45\linewidth]{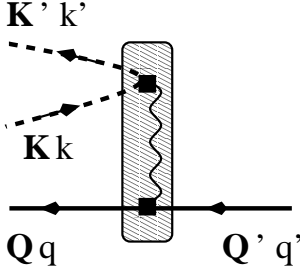}
\caption{Diagrammatic representation of the golden rule~(\ref{GoldenRule}) for a transition
from the surface state $(\vec{Q}'q')$ to the surface state $(\vec{Q}q)$ (solid lines) via
scattering on an internal electron (dashed line) which can be interpreted as the coupling
to internal electron-hole pairs. The wavy line denotes the screened Coulomb interaction
between internal and external electrons and the box symbolizes the dynamic, that is,
inelastic electron-metal interaction.}
\label{GR}
\end{figure}

The Hamiltonian~(\ref{PhysiH}) with $H_e$, $H_s$, and $H_{es}$ respectively given 
by~(\ref{He}), (\ref{Hs}), and (\ref{Hes}) can be used to calculate the transition 
rate from any initial surface state $\psi_{\vec{Q}'q'}$ to any final surface state 
$\psi_{\vec{Q}q}$. For the sticking process the initial state belongs to the continuum
of surface states and the final state is a bound surface state while for the desorption
process it is vice versa. In lowest order perturbation theory (see Fig.~\ref{GR}), the rate 
is given by the golden rule,
\begin{eqnarray}
{\cal W}(\vec{Q}q,\vec{Q}'q')&=&\frac{2\pi}{\hbar}\sum_{\vec{K}\vec{K}'}\sum_{kk'}
\big|V_{\vec{Q}'q'\vec{K}'k'}^{\vec{Q}q~~\vec{K}k}\big|^2
\nonumber\\
&\times& n_F(E_{\vec{K}'k'})[1-n_F(E_{\vec{K}k})]
\nonumber\\
&\times& \delta(E_{\vec{Q}'q'}+E_{\vec{K}'k'}-E_{\vec{Q}q}-E_{\vec{K}k})~,
\label{GoldenRule}
\end{eqnarray}
where $n_F(E)=1/(\exp[(E-E_F)/k_BT_s]+1)$ is the Fermi distribution function for the metal
electrons with Fermi energy $E_F$ and temperature $T_s$. 

To calculate the matrix element~(\ref{Ielement}) we need the solutions of the Schr\"odinger 
equations~(\ref{SGpsi}) and~(\ref{SGphi}). Physisorption of electrons involves transitions 
between bound and unbound surface states. The matrix elements for these transitions are 
large when the spatial overlap between the initial and final states is large. With unbound 
surface states inside the gap, image states, that is, bound surface states close to the zero 
of the energy axis (see Fig.~\ref{SurfaceStates}), have the largest overlap. Crystal-induced 
surface states, having most weight in regions where the weight of unbound surface states is 
very small, give rise to a smaller overlap and are thus less important. 
We neglect therefore crystal-induced surface states and replace 
$V(z)$ in~(\ref{SGpsi}) by
\begin{eqnarray}
V(z)\rightarrow\left\{\begin{array}{ll}
\infty & \mbox{for}~z\le 0\\
V_p(z) & \mbox{for}~z>0~,
\end{array}\right.
\label{Vapprox}
\end{eqnarray}
where 
\begin{eqnarray}
V_p(z)=-\frac{e^2}{4z} 
\label{VimMetal}
\end{eqnarray}
is the classical image potential. As explained in appendix A, $V_p$  can 
be understood in terms of virtual surface plasmon excitations~\cite{RM72,EM73,Barton81}. 
We thus calculated the surface states as if the energy gap on the surface were infinitely
large. The solutions of~(\ref{SGpsi}) are then Whittaker functions which vanish for
$z\le 0$ (see Appendix B) and the required matrix elements can be obtained analytically. 
\begin{figure}[t]
\vspace{5mm}
\includegraphics[width=0.96\linewidth]{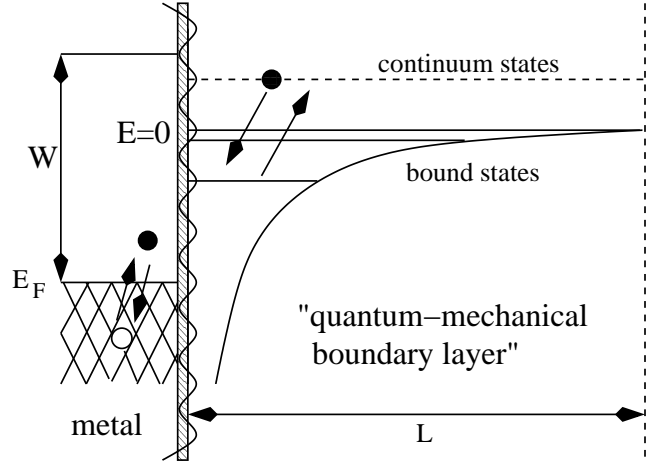}
\caption{Illustration of the microphysics for a plasma electron approaching a
metallic boundary. The wavy line indicates the surface mode
responsible for the attractive polarization potential (\textquotedblleft
image potential\textquotedblright) and $L$ is the width of the \textquotedblleft
boundary layer\textquotedblright~where a quantum-mechanical calculation applies.
The electron looses (gains) energy due creation (annihilation) of electron-hole
pairs in the metal. Due to these processes it may get trapped in (escape from)
the bound states of the polarization potential. In other words, it may get stuck
at (desorb from) the plasma boundary. $W$ and $E_F$ are, respectively, the work
function and the Fermi energy of the metal.}
\label{Jellium}       % Give a unique label
\end{figure}

As far as volume states required for the construction of internal electron-hole
pairs are concerned, we followed~\cite{NNS86,WJS92} and calculated these states as if 
the work function $W$ of the metal where infinite. Measuring moreover energies inside 
the material from the average of the crystal potential and neglecting the oscillations
of the potential, that is, treating the metal boundary as a jellium halfspace,
\begin{eqnarray}
\tilde{V}(z)\rightarrow \left\{\begin{array}{ll}
0 & \mbox{for}~z<0\\
\infty & \mbox{for}~z\ge 0~.
\end{array}\right.
\label{Vtildeapprox}
\end{eqnarray}
The wavefunctions $\phi_k(z)$ vanish then for $z\ge 0$ and are standing waves for $z<0$. 
Using box-normalization,
\begin{eqnarray}
\phi_k(z)=\sqrt{\frac{2}{L}}\sin(kz)
\label{StatesForInternalElectron}
\end{eqnarray}
leading to $\tilde{E}_k=\hbar^2k^2/2m_e$ with $k=\pi n/L$ and $n\ge 1$ an integer. In the final 
expressions for $s_e$ and $\tau_e$ we use $L\rightarrow\infty$ making $k$ continuous. 

The physical content of the simplified planar model is summarized in Fig.~\ref{Jellium}. 
It will be used in the next two subsections to calculate, respectively, $s_e$ and $\tau_e$
for a metallic plasma boundary. Due to the approximate potentials~(\ref{Vapprox}) 
and~(\ref{Vtildeapprox}), external and internal single electron wavefunctions vanish in 
complementary halfspaces. As a result, the matrix element~(\ref{Ielement}) factorizes,
\begin{eqnarray}
I_{q'k'}^{qk}(\vec{Q})=I_{qq'}^{(1)}(\vec{Q})I_{kk'}^{(2)}(\vec{Q})
\label{factorization}
\end{eqnarray}
with
\begin{eqnarray}
I_{qq'}^{(1)}(\vec{Q})\!\!&=&\!\!\int_0^\infty\!\!\!dz\exp[-z\sqrt{k_s^2+\vec{Q}^2}]\psi_q^*(z)\psi_{q'}(z)~,
\label{I(1)}\\
I_{kk'}^{(2)}(\vec{Q})\!\!&=&\!\!\int_0^\infty\!\!\!dz\exp[-z\sqrt{k_s^2+\vec{Q}^2}]\phi_k^*(-z)\phi_{k'}(-z)~
\label{I(2)}
\end{eqnarray}
to be calculated explicitly in appendix B. 

A rigorous calculation of $s_e$ and $\tau_e$, taking for instance into account that sticking 
and desorption occur on different timescales~\cite{Brenig82}, should be based on quantum-kinetic
master equations for the time-dependent occupancies of the surface states $\psi_{\vec{Q}q}$. 
The master equations could be derived from~(\ref{PhysiH}) with techniques from non-equilibrium
physics~\cite{KG86}. In lowest order perturbation theory, the transition rates appearing in the
master equation would be given by~(\ref{GoldenRule}). In the following, we will not use this
advanced approach. Instead we will calculate $s_e$ and $\tau_e$ perturbatively by appropriately 
summing and weighting the transition rate~(\ref{GoldenRule}) over initial and final states. 

\subsection{Sticking coefficient}
\label{SectionStickCoeff}

In order to calculate the sticking coefficient $s_e$ we consider the positive half 
space ($z>0$) as a kind of quantum-mechanical boundary layer (see Fig.~\ref{Jellium}). 
A measure of the tendency $S_{\vec{Q}n,\vec{Q}'q'}$ with which an electron approaching in
an unbound state $\Psi_{\vec{Q}'q'}$ the plasma-boundary at $z=0$ gets stuck in a bound state 
$\Psi_{\vec{Q}n}$ is then the time it takes the electron to traverse the boundary layer 
forwards and backwards divided by the time it takes the electron to make a transition from 
$\Psi_{\vec{Q}'q'}$ to $\Psi_{\vec{Q}n}$~\cite{GS91}. 

Since the width of the quantum-mechanical boundary layer is $L$, 
\begin{eqnarray}
S_{\vec{Q}n,\vec{Q}'q'}=\frac{2L}
{_{\rm in}\langle\vec{Q}'q'|\frac{\vec{p}\cdot\vec{n}}{m_e}|\vec{Q}'q'\rangle_{\rm in}}
\times\frac{1}{{\cal W}^{-1}(\vec{Q}n,\vec{Q}'q')}~,
\label{sQQ}
\end{eqnarray}
where the denominator in the first factor is the velocity matrix element calculated with 
the incoming part of the state $\Psi_{\vec{Q}'q'}$; $\vec{n}$ is the normal vector
of the boundary pointing towards the plasma and $\vec{p}=-i\hbar\nabla$, the 
quantum-mechanical momentum operator. Using the asymptotic form of the unbounded 
wavefunctions given in Eq.~(\ref{AsymptExp}) of appendix B, we find 
\begin{eqnarray}
_{\rm in}\langle\vec{Q}'q'|\frac{\vec{p}\cdot\vec{n}}{m_e}|\vec{Q}'q'\rangle_{\rm in}
=\frac{\hbar q'}{8m_e a_B}~.
\end{eqnarray}
Hence, 
\begin{eqnarray}
S_{\vec{Q}n,\vec{Q}'q'}=\frac{16 L m_e a_B}{\hbar q'}{\cal W}(\vec{Q}n,\vec{Q}'q')~.
\end{eqnarray}
The tendency with which the electron approaching the boundary in the state 
$\Psi_{\vec{Q}'q'}$ gets stuck in any one of the bound states -- the energy 
resolved sticking coefficient -- is then simply given by
\begin{eqnarray}
S_{\vec{Q}'q'}&=&\sum_{\vec{Q}n}S_{\vec{Q}n,\vec{Q}'q'}
\nonumber\\
&=&\frac{16L m_e a_B}{\hbar q'}\sum_{n\vec{Q}}{\cal W}(\vec{Q}n,\vec{Q}'q')~.
\label{sQ}
\end{eqnarray}

The sticking coefficient $s_e$ entering the rate equation~(\ref{REqe}) is an
energy-averaged sticking coefficient resulting from an appropriately performed sum 
over $S_{\vec{Q}'q'}$. As mentioned before, a rigorous derivation of an expression
for $s_e$ should be based on the master equation for the occupancies of the surface 
states~\cite{Brenig82,KG86}. A simpler way to obtain $s_e$ is however to regard the wall 
as a particle detector. The global sticking coefficient can then be defined as
\begin{eqnarray}
\sum_{\vec{Q}'q'}S_{\vec{Q}'q'}q'n_{\vec{Q}'q'}
=s_e\sum_{\vec{Q}'q'}q'n_{\vec{Q}'q'}~,
\label{se}
\end{eqnarray}
where $n_{\vec{Q}'q'}$ are the occupancies of the unbound surface states $\Psi_{\vec{Q}'q'}$. 

The occupancies $n_{\vec{Q}'q'}$ depend on the properties of the plasma. It is 
tempting to simply identify $n_{\vec{Q}'q'}$ with the incoming part of the electron 
distribution function as it arises on the surface from the solution of the 
Boltzmann-Poisson equations. However, one should keep in mind that the distribution 
function is a classical object whereas $n_{\vec{Q}'q'}$ is a quantum-mechanical expectation 
value. There arises therefore the question how the quantum-mechanical processes 
encoded in the above equations can be properly fed into the semiclassical description
of the plasma in terms of Boltzmann-Poisson equations. The issue is subtle because at 
the plasma boundary the potential varies so rapidly that the basic assumptions of the 
validity of the Boltzmann equation no longer hold. Mathematically, the microphysics 
should be put into a surface scattering kernel, course-grained over a few $nm$, which 
connects, generally retarded in time, the incoming electron distribution function with 
the outgoing one. But even for neutral particles, a microscopic derivation of such a 
scattering kernel has not yet been given. There exist only more or less plausible 
phenomenological expressions which parameterize the kernel with 
accommodation coefficients~\cite{Kuscer71}. 

From the boundary-layer point of view used in the derivation of Eqs.~(\ref{sQQ})--(\ref{se}),
the plasma, or, more precisely, the sheath of the plasma, is infinitely far away from the 
plasma boundary. Rigorously speaking, we can thus say nothing about how the microphysics 
at the plasma boundary merges with the physics in the plasma sheath. 

To make nevertheless
contact with the plasma we have to guess how the unbound surface states $\Psi_{\vec{Q}'q'}$ 
are occupied. For simplicity we assume Maxwellian occupancy, with an electron temperature
$T_e=(k_B\beta_e)^{-1}$, but other guesses, more appropriate for the plasma 
sheath, are also conceivable. For Maxwellian electrons, the global sticking coefficient
is given by
\begin{eqnarray}
s_e=\frac{\sum_{\vec{Q}'q'} S_{\vec{Q}'q'} q'
\exp[-\beta_e E_{\vec{Q}'q'}]}{\sum_{\vec{Q}'q'} q' \exp[-\beta_e E_{\vec{Q}'q'}]}~.
\label{seSum}
\end{eqnarray}

In the limit $L\rightarrow\infty$ and $A\rightarrow\infty$ the momentum summations in 
Eqs.~(\ref{sQQ})--(\ref{seSum}) become integrals. The calculation of $S_{\vec{Q}'q'}$ and
$s_e$ reduces therefore to the calculation of high-dimensional integrals. In appendix 
C we describe the approximations invoked for the integrals. Some of the integrals can
then be analytically performed. But the final expressions for the sticking coefficients 
remain multi-dimensional integrals, which have to be done numerically.

Measuring energies in units of $R_0$ and distances in units of $a_B$, Eq.~(\ref{seSum}) 
for the global sticking coefficient reduces to 
\begin{eqnarray}
s_e=\bigg(\frac{4}{\pi}\bigg)^2\frac{\beta_e^{3/2}}{\beta_s^{1/2}}
I_{\rm stick}~,
\label{seFinal}
\end{eqnarray}
where 
\begin{eqnarray}
I_{\rm stick}=\int_0^\infty\!\!\!dR\int_{-\infty}^\infty\!\!\!d\omega 
\frac{1+n_B(\omega)}{1+(R/k_s)^2}
h(R,\omega)g(R,\omega)
\label{Istick}
\end{eqnarray}
with $n_B(E)=1/(\exp[\beta_sE]-1)$ the Bose distribution function and 
$h(R,\omega)$ and $g(R,\omega)$ two functions defined, respectively, 
%the electronic matrix elements~(\ref{I(1)}) and~(\ref{I(2)}). They are 
%defined, respectively, 
in appendix C by Eq.~(\ref{hfct}) and~(\ref{gfct}).
\begin{figure}[t]
\vspace{5mm}
\includegraphics[width=0.96\linewidth]{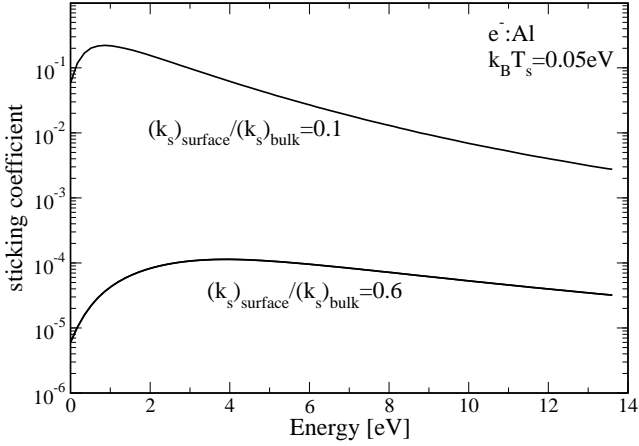}
\caption{Energy resolved sticking coefficient for an electron hitting
perpendicularly an aluminum surface at $k_BT_s=0.05eV$. The
screening wavenumber for the Coulomb interaction between an incident plasma
electron and an internal aluminum electron, $(k_s)_{\rm surface}$, is not
well known. Results are therefore shown for
$(k_s)_{\rm surface}/(k_s)_{\rm bulk}=0.1$ (weak screening, strong coupling)
and for $(k_s)_{\rm surface}/(k_s)_{\rm bulk}=0.6$ (moderate screening, weak
coupling); $(k_s)_{\rm bulk}$ is the screening wavenumber of aluminum (see
table~\ref{ScreenPara}). Since $(k_s)_{\rm surface}=0.6(k_s)_{\rm bulk}$ is
most probably the relevant screening parameter~\cite{NNS86,WJS92}, the sticking
coefficient is rather small.}
\label{Figsperp}       % Give a unique label
\end{figure}

Below we also present results for the energy resolved sticking coefficient for 
perpendicular incidence $(\vec{Q}'=0)$. It is given by 
\begin{eqnarray}
S^\perp_{E'}=\bigg(\frac{4}{\pi}\bigg)^2\!\frac{\pi^{1/2}}{\beta_s^{1/2}}
\int_0^\infty\!\!\!\!dR g^\perp(R,E')~,
\label{sperp}
\end{eqnarray}
with $E'=q'^2$ and $g^\perp(R,E')$ a function defined in appendix C,
Eq.~(\ref{gperp}).

The functions $h(R,\omega)$, $g(R,\omega)$, and $g^\perp(R,E')$ contain
summations over the Rydberg series of bound surface states. If not stated 
otherwise, we truncated these sums after $N=15$ terms. These functions are 
moreover defined in terms of integrals which can be done only numerically.
We use Gaussian integration with $40-80$ integration points. More specifically,
$h(R,\omega)$ and $g^\perp(R,E')$ are one-dimensional integrals while 
$g(R,\omega)$ is a two-dimensional one. Hence, $s_e$ and $S^\perp_{E'}$ are
given by a five-dimensional and a two-dimensional integral, respectively. 
\begin{figure}[t]
\vspace{5mm}
\includegraphics[width=0.96\linewidth]{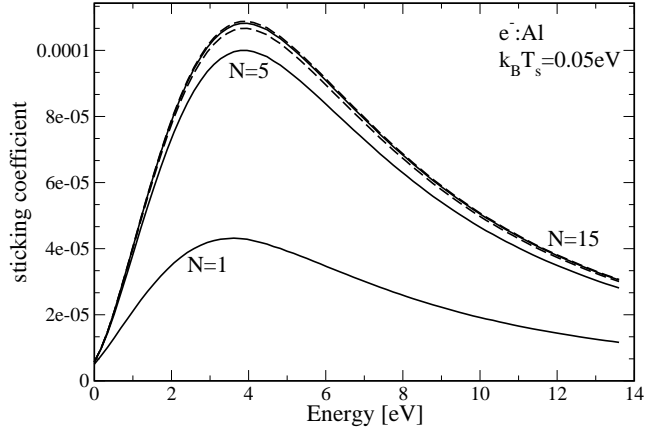}
\caption{Dependence of the energy resolved sticking coefficient for perpendicular
incidence on the number $N$ of bound states included in the calculation. Except of
the screening wavenumber, which is set to $(k_s)_{\rm surface}=0.6(k_s)_{\rm bulk}$,
the parameters are identical to the ones used in Fig.~\ref{Figsperp}. The dashed
lines are for $N=10$ and $N=20$, respectively, indicating the fast converge with
respect to $N$.
}
\label{FigsperpN}       % Give a unique label
\end{figure}

In the formulae for the sticking coefficients we multiplied the binding energies 
of the surface states $|E_n|$ obtained from Eq.~(\ref{SGpsi}) by an overall 
factor of $0.7$. This value was chosen to bring the binding energy of the lowest 
surface state $|E_1|=0.85eV$ in accordance with the experimentally measured value 
for copper: $|E_1|^{\rm Cu}\approx0.6eV$~\cite{EL95}. For other metals we used the
same correction factor. 

Figure~\ref{Figsperp} shows the results for $S^\perp_{E'}$ when an electron with 
energy $E'$ hits perpendicularly an aluminum boundary at $k_BT_s=0.05eV$. 
Representative for weak and moderate screening we plotted data for 
$(k_s)_{\rm surface}/(k_s)_{\rm bulk}=0.1$ and $(k_s)_{\rm surface}/(k_s)_{\rm bulk}=0.6$.
The latter is the screening parameter used in~\cite{NNS86,WJS92} 
to study the interaction of positrons with an aluminum surface. If the corresponding
value for $1/(k_s)_{\rm surface}$ is indeed a reasonable estimate for the length on which
the Coulomb interaction between an external and an internal electron is screened, the 
sticking coefficient for electrons should be extremely small, of the order of $10^{-4}$. 
Only for weak screening, and thus strong coupling, does $S^\perp_{E'}$ approach values 
of the order of $10^{-1}$ which are perhaps closer to the value one would expect on first
sight. 

To clarify the contribution the various bound states have to the 
sticking coefficient, we plot in Fig.~\ref{FigsperpN} the dependence of 
$S_{E'}^\perp$ on the number $N$ of bound states included in the 
calculation. As can be seen, the lowest bound state ($N=1$) contributes only
roughly $40\%$ to the total $S_{E'}^\perp$. The sticking coefficient
increases then with increasing $N$ but converges for $N\approx 10-20$. Because of
this fast convergence we present all results below only for $N=15$. The reason 
for the convergence can be traced back to the decrease of the electronic matrix 
element $I^{(1)}_{kn}(\vec{Q})$ defined in Eq.~(\ref{I(1)}), which we approximate 
by $I^{(1)}_{k\ll 1n}(\vec{Q}=0)$ (see appendix C), with 
increasing $n$, where $n=1,2,...$ labels the bound surface states.

\begin{figure}[t]
\vspace{5mm}
\includegraphics[width=0.96\linewidth]{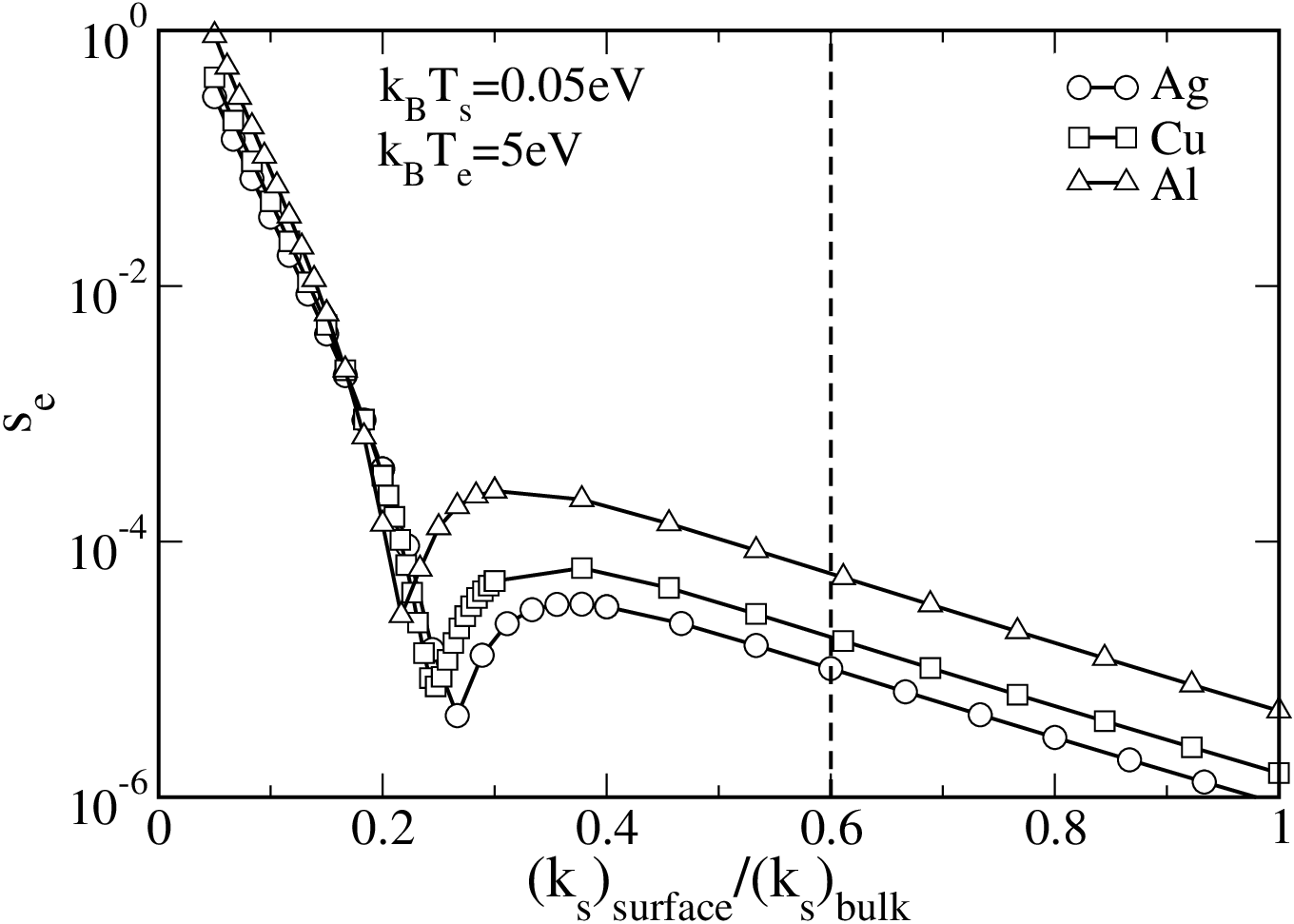}
\caption{The global sticking coefficient $s_e$ for a thermal beam of electrons
with $k_BT_e=5eV$ hitting various metal surfaces at $k_BT_s=0.05eV$ as a function of
$(k_s)_{\rm surface}/(k_s)_{\rm bulk}$, where $(k_s)_{\rm bulk}$ is the
screening wavenumber in the bulk of the respective metal (see table~\ref{ScreenPara}).
Following~\cite{NNS86,WJS92}, we would expect $(k_s)_{\rm surface}=0.6(k_s)_{\rm bulk}$
to be a reasonable estimate for the screening parameter. Hence, $s_e\approx 10^{-5}-10^{-4}$.}
\label{sgl_f}       % Give a unique label
\end{figure}

Global sticking coefficients $s_e$ as a function of the screening wavenumber 
$(k_s)_{\rm surface}$ are shown in Fig.~\ref{sgl_f} for different metals. For
$(k_s)_{\rm surface}/(k_s)_{\rm bulk}>0.4$, 
the sticking coefficients are again extremely small. As 
expected they increase with decreasing $(k_s)_{\rm surface}/(k_s)_{\rm bulk}$,
reaching values close to unity for weak screening. In this strong coupling 
regime, our perturbative calculation of $s_e$ is no longer valid. We believe
however that $(k_s)_{\rm surface}/(k_s)_{\rm bulk}<0.4$ is unphysical. 
The kink around $(k_s)_{\rm surface}/(k_s)_{\rm bulk}\approx 0.25$ must
be due to an accidental resonance in $g(R,\omega)$. It is of no physical
significance. 

Why is the sticking coefficient for electrons so small? We have no satisfying 
explanation. Our calculation produces small a sticking coefficient because the 
matrix element~(\ref{Ielement}) turns out to be very small. We certainly 
underestimate it because the wavefunctions of 
the approximate potentials~(\ref{Vapprox}) and~\ref{Vtildeapprox}) vanish in
complementary halfspaces, in contrast to the exact wavefunctions which have 
tails. Nevertheless it is hard to image the tails of the wavefunctions to increase
the matrix elements by three orders of magnitude. 

The approximations we had to make to end up with manageable equations for $s_e$, 
in particular, the assumptions about the momentum dependence of the electronic matrix 
elements (see appendix C and, for a discussion, the next section) should also not 
lead to a sticking coefficient which is more than one order of magnitude off. In 
this respect let us emphasize that in contrast to the calculations performed 
in~\cite{NNS86,WJS92} for a positron, which produce positron sticking coefficients 
of the order of $0.1$, we use the eigenenergies and eigenstates of the $1/z$ 
potential and not the ones of an artificial box potential. 

Usually it is assumed that $s_e$ is also at least of the order of $0.1$~\cite{DS87}. 
This expectation seems to be primarily based on the semiclassical back-on-the 
envelop-estimate of Umebayashi and Nakano~\cite{UN80}. It is thus appropriate to discuss
their approach in some detail.

From the energy $\Delta E_s$ an electron can exchange in a single classical 
collision with the constituents of the solid they first estimated, using the 
analogy to the M\"ossbauer effect, the probability $\alpha$ for inelastic 
one-phonon emission. For that purpose, they had to estimate the number $N_c$ of
constituents of the surface an electron with a de-Broglie wavelength 
corresponding to its kinetic energy $E_0$, $\lambda_e^{dB}=2\pi a_B\sqrt{R_0/E_0}$, 
simultaneously impacts. A rough estimate is $N_c=(\lambda_e^{dB}/a)^2$, where $a$ 
is the lattice constant of the material. Under the assumption that the 
electron hops along the surface they then calculated the probability with 
which the electron does not escape after $l$ hops where $l$ is the number
of inelastic collisions which are necessary for the electron to transfer its 
whole positive kinetic energy to the lattice, that is, to end up in a state of 
negative energy. Identifying this probability with the (global) sticking 
coefficient, they obtained
\begin{eqnarray}
s_e=\prod_{i=0}^{l-1}\frac{1}{1+\beta_i/\alpha}~,
\label{UN}
\end{eqnarray}
where $\beta_i=(E_0-i\Delta E)/E_b$ is the escape probability after $i$ inelastic
collisions~\cite{HS70}, $\Delta E=2\Delta E_s/3N_c\alpha$, $\Delta E_s=4m_e(E_0+E_b)/M$, 
$E_b$ is the depth of the surface potential and $M$ is the mass of the 
constituents of the solid.

\begin{figure}[t]
\vspace{5mm}
\includegraphics[width=0.96\linewidth]{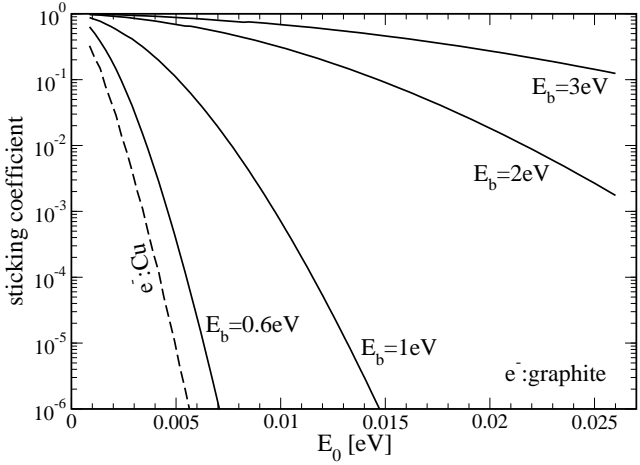}
\caption{Electron sticking coefficient obtained from Umebayashi and Nakano's
phenomenological model~\cite{UN80}, see Eq.~(\ref{UN}). The solid lines are for the
$e^{-}$:graphite system originally considered by them ($k_BT_D=420K$,
$M_{\rm C}=12m_p$, where $m_p$ is the proton mass, and $a=2.5\AA$)
and the dashed line is for an $e^{-}$:Cu system ($E_b=0.6eV$, $k_BT_D=343K$,
$M_{\rm Cu}=64m_p$, and $a=3.61\AA$). The sticking coefficient diminishes
rapidly with increasing electron energy and approaches one at zero electron energy,
in contrast to what one would expect from a quantum-mechanical calculation~\cite{BR92}.}
\label{sgl_UN}       % Give a unique label
\end{figure}

Sticking coefficients for graphite obtained from Eq.~(\ref{UN}) are shown in Fig.~\ref{sgl_UN}. 
Within Umebayashi and Nakano's semiclassical approach we identified $E_b$ with the binding
energy of the electron. According to Fig.~\ref{sgl_UN} the sticking coefficient very 
quickly approaches extremely small values with increasing energy $E_0$. The smaller the 
binding energy $E_b$, the faster the decrease. The values for $s_e$ originally given by Umebayashi
and Nakano were for kinetic energies smaller than $0.0026eV$ and binding energies larger than 
$1eV$. Only in this parameter regime is the sticking coefficient close to one. In the 
parameter range which is of interest to us (kinetic and binding energies at least a few 
tenth of an electron volt) Umebayashi and Nakano's estimate gives also an extremely small 
sticking coefficient. 

We should of course not directly compare the results obtained from Eq.~(\ref{seFinal}) with 
the ones obtained from Eq.~(\ref{UN}) because Eq.~(\ref{seFinal}) assumes energy relaxation 
due to internal electron-hole pairs whereas Eq.~(\ref{UN}) assumes energy relaxation due to
phonons. However, a quantum-mechanical calculation of the phonon-induced electron sticking 
coefficient at vanishing lattice temperature also shows that $s_e\approx 10^{-4}$~\cite{BR92}, 
in contrast to what Umebayashi and Nakano find. Although they incorporate some quantum mechanics
their approach is basically classical. It is based on the notion of a classical particle hopping
around on the surface and exchanging energy with the solid in binary encounters. As in any 
classical theory for the sticking coefficient, it is therefore not surprising that the sticking
coefficient they obtain approaches unity for the low energies they consider~\cite{KG86}. 

\subsection{Desorption time}
\label{SectionDesorbTime}

We now calculate the electron desorption time $\tau_e$. For that purpose, we have to 
specify the occupancies of the bound electron surface states. In general, this is a 
critical issue. However, provided the desorption time $\tau_e$ is much larger than the 
time it takes to establish thermal equilibrium with the boundary, it is plausible to 
assume that bound electron surface states are populated according to
\begin{eqnarray}
n_{\vec{Q}n}\sim\exp[-\beta_sE_{\vec{Q}n}]~, 
\label{nequilibrium}
\end{eqnarray}
where $T_s=1/k_B\beta_s$ is the surface temperature. 

Desorption is accomplished as soon as the electron is in any one of the unbound surface 
states. Hence, the inverse of the desorption time, that is, the desorption rate, is 
given by~\cite{KG86}
\begin{eqnarray}
\frac{1}{\tau_e}=
\frac{\sum_{\vec{Q}'n'}\sum_{\vec{Q}q}\exp[-\beta_sE_{\vec{Q}'n'}]{\cal W}(\vec{Q}q,\vec{Q}'n')}
{\sum_{\vec{Q}n}\exp[-\beta_sE_{\vec{Q}n}]}~,
\end{eqnarray}
where ${\cal W}(\vec{Q}q,\vec{Q}'n')$ is the transition rate from the bound surface 
state $(\vec{Q}',n')$ to the unbound surface state $(\vec{Q},q)$ as defined by the 
golden rule~(\ref{GoldenRule}).
\begin{figure}[t]
\vspace{5mm}
\includegraphics[width=0.96\linewidth]{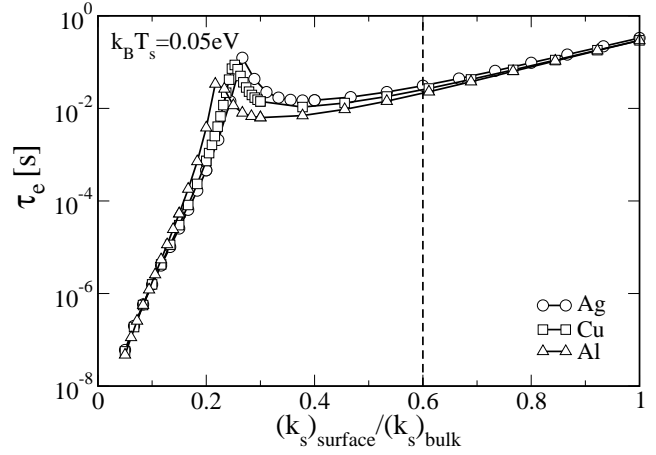}
\caption{Desorption time $\tau_e$ for an electron bound in the polarization-induced
external surface states of various metal surfaces at $k_BT_s=0.05eV$ as a function
of $(k_s)_{\rm surface}/(k_s)_{\rm bulk}$. Since $(k_s)_{\rm surface}=0.6(k_s)_{\rm bulk}$
is most probably the relevant screening wavenumber~\cite{NNS86,WJS92},
$\tau_e\approx 10^{-2}s$. For the used material parameters, see table~\ref{ScreenPara}.}
\label{tau_f}       % Give a unique label
\end{figure}

Measuring again energies in units of $R_0$ and distances in units of $a_B$ and using the 
same approximations as in the calculation of the sticking coefficient (see appendix C) 
the desorption rate can be cast into 
\begin{eqnarray}
\tau_e^{-1}
=\frac{R_0}{2\pi^3\hbar Z}I_{\rm desorb}~,
\label{taue}
\end{eqnarray}
where
\begin{eqnarray}
I_{\rm desorb}&=&\int_0^\infty\!\!\!\!dR\!\int_{-\infty}^\infty\!\!\!\!\!d\omega
\frac{1+n_B(\omega)}{1+(R/k_s)^2}
f(R,\omega)g(R,\omega)~,
\label{Idesorb}
\end{eqnarray}
$Z=\sum_n\exp[-\beta_sE_n]$, $n_B(E)$ is again the Bose distribution function, and 
$f(R,\omega)$ is an one-dimensional integral defined in appendix C, Eq.~(\ref{ffct}).
Thus, to obtain $\tau_e^{-1}$ from Eq.~(\ref{taue}) we have 
to do a five-dimensional integral. As for the calculation of $s_e$ we again use Gaussian 
quadratures for that purpose.

In Figure~\ref{tau_f} we present, as a function of the screening parameter,
numerical results for $\tau_e$ for an electron bound in the polarization-induced external 
surface states of various metal surfaces at $k_BT_s=0.05eV$. To be close to reality, we 
again corrected the binding energies $|E_n|$ by a factor $0.7$. As can be seen, 
except for small screening parameters and thus strong coupling, $\tau_e\approx 10^{-2}s$. 

Compared to typical desorption times for neutral mole- cules, which are of the order of $10^{-6}s$
or less~\cite{KG86}, the electron desorption time we find is rather long. This is a consequence 
of our assumption that the bound electron is in thermal equilibrium with the surface (viz: 
Eq.~(\ref{nequilibrium})) and the fact that the binding energy of the lowest surface state 
$|E_1|\gg k_BT_s$. Thus, the electron desorbs de facto from the lowest surface state which 
has a binding energy of $\sim 0.6eV$. The binding energies for neutral molecules, on the other
hand, are typically one order smaller and thus of the order of $k_BT_s$ resulting in much larger
desorption rates and thus shorter desorption times. 

In the model for the quasi-stationary charge of a dust particle presented in the previous 
section the product $(s\tau)_e$ was of central importance. Combining~(\ref{seFinal}) 
and~(\ref{taue}), the microscopic approach gives 
\begin{eqnarray}
(s\tau)_e=\frac{h}{(k_BT_e)^{3/2}(k_BT_s)^{-1/2}}
\frac{16 I_{\rm stick}}{I_{\rm desorb}}\sum_n\exp[\beta_s|E_n|]~,
\label{stauFull}
\end{eqnarray}
where $I_{\rm stick}$ and $I_{\rm desorb}$ are defined in Eqs.~(\ref{Istick}) and~(\ref{Idesorb}), 
respectively. 
\begin{figure}[t]
\vspace{5mm}
\includegraphics[width=0.96\linewidth]{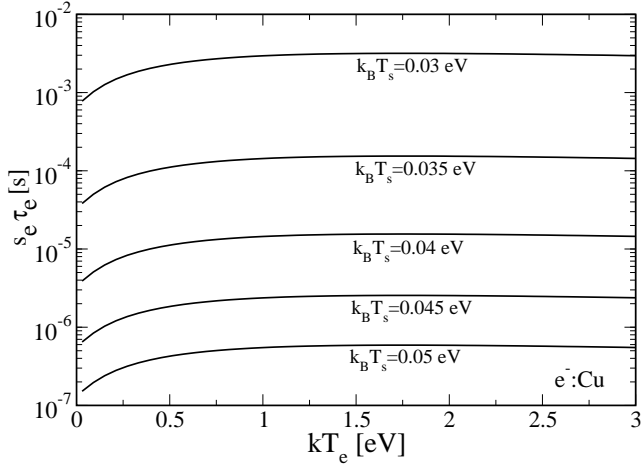}
\caption{The product $s_e\tau_e$ as a function of $k_BT_e$ for a thermal beam of
electrons hitting a copper surface at various temperatures $k_BT_s$. The surface
screening wavenumber is set to $(k_s)_{\rm surface}=0.6(k_s)_{\rm bulk}$, where
$(k_s)_{\rm bulk}$ is the screening wavenumber for copper (see table~\ref{ScreenPara}).}
\label{stau_Ts}       % Give a unique label
\end{figure}

Figure~\ref{stau_Ts} shows numerical results for $(s\tau)_e$ for a copper 
surface as a function of the electron and surface temperature. The screening wavenumber
$(k_s)_{\rm surface}=0.6(k_s)_{\rm bulk}$ and the binding energies are again corrected
by the factor $0.7$ which makes $|E_1|$ to coincide with the experimental value for 
copper. Notice, the weak dependence of the product $(s\tau)_e$ on the electron temperature 
and the rather strong dependence on the surface temperature. The latter is of course a 
consequence of the exponential function in Eq.~(\ref{stauFull}). 
Although the sticking coefficient and desorption times have values which are perhaps in 
contradiction to naive expectations, $s_e$ being extremely small and $\tau_e$ being rather 
large, the product $(s\tau)_e$ has the order of magnitude expected from our surface 
model (see section 2). In particular, $(s\tau)_e\simeq 10^{-6}s$ for $k_BT_s=0.045eV$ would 
produce grain charges of the correct order of magnitude. Thus, using Eq.~(\ref{stauFull}) 
instead of Eq.~(\ref{stau}) and $k_BT_s$ as an adjustable parameter, which is still necessary 
because the grain temperature is unknown, we could produce, for physically realistic surface 
temperatures, surface charges for metallic grains which are in accordance with 
experiment~\cite{WHR95}. 

Although the microscopic Eq.~(\ref{stauFull}) has a similar structure as the phenomenological 
expression~(\ref{stau}) there are significant differences. First, the microscopic formula 
contains more than one bound state and depends not only on $T_s$ but also on $T_e$. In 
addition, there is a numerical factor $16=2\times 8$ where the factor $2$ comes from the fact 
that an electron traversing the quantum-mechanical boundary layer can make a transition to a
bound state on its way towards the surface and on its way back to the plasma and the factor $8$
originates from the asymptotic form of the wavefunction for the incoming electron. The 
phenomenological approach simply assumes here a plane wave whereas the microscopic approach works
with Whittaker functions (see appendix B and C). Most importantly, however, the two functions 
$I_{\rm stick}$ and $I_{\rm desorb}$, which depend on the microscopic details of the inelastic 
scattering process driving physisorption, and thus on the electron and surface temperature as 
well as material parameters such as the screening wavenumber, are in general not identical.
Hence, $I_{\rm stick}/I_{\rm desorb}\neq 1$.

For the hypothetical case of a single bound state, however, whose binding energy $|E_1|$ is 
much larger than $k_BT_s$ and $k_BT_e$, Eq.~(\ref{stauFull}) reduces to a form which, for 
$k_BT_e=k_BT_s$, becomes identical to the phenomenological expression~(\ref{stau}), except 
of the numerical factor referred to in the previous paragraph. The simplification arises
because for low temperatures the integrals defining $I_{\rm stick}$ and $I_{\rm desorb}$ 
can be calculated asymptotically within Laplace's approximation (see appendix C). The 
sticking coefficient and desorption time are then given by 
\begin{eqnarray}
s^L_e&=&\frac{4\big|I_1^{(1)}\big|^2\bar{g}}{\pi\beta^{1/2}_s\beta^{1/2}_e}~, 
\label{seL}\\
\tau^L_e&=&8\pi^2\beta_s^2\frac{\hbar}{R_0}\frac{\exp[\beta_s|E_1|]}{\big|I^{(1)}_1\big|^2\bar{g}}~,
\label{taueL}
\end{eqnarray}
with $\bar{g}$ defined in appendix C, Eq.~(\ref{gbar}). Thus, the product,
\begin{eqnarray}
(s\tau)_e^L=\frac{16h}{(k_BT_s)^{3/2}(k_BT_e)^{-1/2}}\exp[\beta_s|E_1|]~,
\label{stauL}
\end{eqnarray}
is independent of the microscopic details of the inelastic scattering processes encoded in
the product $|I^{(1)}_1|^2\bar{g}$. Identifying $|E_1|$ with the electron desorption energy 
$E_e^d$ and setting $k_BT_e=k_BT_s$, we finally obtain, except of the numerical factor 
$16=8\times 2$, from Eq.~(\ref{stauL}) the  phenomenological expression~(\ref{stau}).

Using Eqs.~(\ref{seL})--(\ref{stauL}) we find for an electron at a copper boundary  
with $k_BT_e=k_BT_s=0.045eV$, $s_e^L=6.23\times 10^{-6}$, $\tau_e^L=0.131s$, and 
$(s\tau)_e^L=8.17\times 10^{-7}s$. Taking only one bound state into account, 
the corresponding values obtained from Eqs.~(\ref{seFinal}) and ~(\ref{taue}) are 
$s_e=4.42\times 10^{-6}$ and $\tau_e=0.135s$, which leads to $(s\tau)_e=6\times 10^{-7}s$,
indicating that at low temperatures Laplace's approximation works indeed reasonably 
well. Since $\tau_e$ does not depend on $k_BT_e$ and $k_BT_s$ is usually much smaller than 
$|E_1|$, approximation~(\ref{taueL}) for $\tau_e$ can be actually always applied, provided
the assumption is correct, that the electron is initially in thermal equilibrium with the 
surface and hence basically in its lowest bound state. The approximation~(\ref{seL}) for 
$s_e$, on the other hand, deteriorates quickly with increasing electron temperature, as 
does the approximation~(\ref{stauL}) for $(s\tau)_e$. 

It is reassuring to be able to derive, under certain conditions and except of a numerical 
factor, whose origin is however clear, from the microscopic expressions for $s_e$ and $\tau_e$ 
the phenomenological relation~(\ref{stau}) we used
in~\cite{BFK08} as an estimate for $(s\tau)_e$. That~(\ref{stauFull}) can be reduced 
to~(\ref{stau}) is a consequence of the perturbative calculation of $s_e$ and $\tau_e$ using 
the golden rule transition rate~(\ref{GoldenRule}) which obeys detailed balance. In this 
respect, our calculation is on par with Lennard-Jones and Devonshire's original microscopic
derivation of the product $s\tau$ for a neutral adsorbant~\cite{LJD36}. In contrast to them, 
we keep however the lateral motion of the adsorbing particle and the elementary excitations
of the solid responsible for energy relaxation are electron-hole pairs and not phonons. 

\subsection{Critique}

In the previous subsections we demonstrated for a particular case, a metallic boundary
with electron energy relaxation due to creation and annihilation of internal 
electron-hole pairs, how a quantum-mechanical calculation can be set up to obtain 
$s_e$ and $\tau_e$ from a microscopic model for the electron-wall interaction. To obtain
manageable equations we had to make various approximations, some were purely technical, 
but others concerned the physics. We now re-state and criticize the approximations
in the hope that it will be read as a list of to-do's. 

We start with the purely technical approximations. In the calculation of the transition 
rate we neglected the dependence of the matrix element~(\ref{Ielement}) on the lateral
momentum transfer and approximated furthermore $I^{(1)}_{kn}(\vec{Q}=0)$ by its leading 
term for $k\ll 1$. Both approximations can be avoided but the final equations become 
more complex and the costs for their numerical handling accordingly higher. At the present 
stage of the investigation this seemed to us not justified. Even more so because we do 
not believe that these approximations are the cause for the unexpectedly small 
values for $s_e$ and the unexpectedly large values for $\tau_e$. Neglecting the 
dependence on the lateral momentum overestimates even the matrix elements, hence 
the transition rate, and thus, eventually, $s_e$ and
$\tau_e^{-1}$. The $k-$dependence of $I_{nk}^{(1)}$ (see Eq.~(\ref{Inq1})
in appendix B), on the other hand, can also not be so large that it increases 
the transition rate by three orders of magnitude as it would be required to 
obtain $s_e\sim 0.1-1$ and $\tau_e\sim 10^{-5}-10^{-6}s$, the values one would 
perhaps naively expect. 

More critical for the matrix element~(\ref{Ielement}) are the replacements~(\ref{Vapprox})
and~(\ref{Vtildeapprox}) because they lead to wavefunctions vanishing in complementary 
halfspaces and thus to the factorization~(\ref{factorization}) of the matrix element. In 
reality the wavefunctions have tails in the complementary halfspaces. A model neglecting 
the tails underestimates therefore the matrix element. In addition, the replacements lead
to the loss of crystal-induced surface states, about which we have more to say below, and 
hard-wire the artificial treatment of surface and volume electrons as two separate species. 
A more realistic modeling should therefore avoid these two approximations.

Both the electron sticking coefficient $s_e$ and the electron desorption time $\tau_e$ 
were obtained from the golden rule for transitions between bound and unbound surface states.
This is only justified for weak coupling and when one quanta of elementary excitation suffices
for the transition. When the coupling is strong, or when more than one quanta are necessary, 
a generalized golden rule has to be used in which the interaction matrix element is replaced 
by the corresponding on-shell T-matrix~\cite{BY73,KG86}. The calculation becomes more 
tedious but it can be done. A principle shortcoming, however, of any approach which uses 
golden-rule-type transition rates directly to
calculate $s_e$ and $\tau_e$ is that it assumes the occupancies of surface states to be only 
weakly affected by the transitions itself. From the physisorption of neutral particles it is
known that this is in general not true~\cite{KT81}.

The calculation of the desorption time, for instance, was based on the assumption that the
desorbing electron is initially in thermal equilibrium with the surface and that during
the desorption process the equilibrium occupancy of the surface states does not change. The 
desorption time is thus much larger than the timescale on which thermal equilibrium at the 
surface is established, in which case the electron basically always desorbs from the lowest 
bound surface state. The equilibration on the surface is controlled by transitions between 
bound surface states. They have to be much faster than transitions between bound and unbound 
surface states. In the golden rule approach this information is put in by hand. Thus, 
although the $\tau_e$ obtained is consistent with the equilibrium assumption, it does not 
justify it. For that purpose, the calculation of $\tau_e$ has to be based on 
quantum-kinetic master equations which include not only transitions between bound and 
unbound surface states but also transitions between two bound surface states~\cite{Brenig82,KG86}.

Master equations are also required when the elementary excitations of the solid have not
enough energy to couple the lowest bound surface states to the continuum. In that case, 
the cascade model developed by Gortel and coworkers~\cite{GKT80a} has to be used. Its 
main idea is that an electron initially bound in a deep state can successively climb up 
to the continuum using weaker bound states as intermediaries. By necessity, it thus 
also contains transitions between bound surface states. 

For metals internal electron-hole pairs provide the most efficient electron energy 
relaxation channel, with phonons and other elementary excitations being unimportant, 
because their energy is either too high (plasmons) or too low (phonons). Both leads to 
severe restrictions in the available phase space. For dielectric boundaries, however, 
it is the energy of internal electron-hole pairs, whose energy is of the order of the 
intrinsic energy gap, which is too high for having any effect. Electron energy relaxation
should then be primarily driven by phonons. Their energy, however, is in the cases of 
interest, for instance, graphite or silicon, too low for promoting an electron from the 
lowest surface states all the way up to the continuum. Hence, $s_e$ and $\tau_e$ have 
to be calculated from Gortel {\it et al.}'s cascade model. When the energy of the phonon 
is moreover not enough to connect two neighboring bound states, the transition rates 
entering the master equation have to be obtained from the generalized golden rule 
containing the T-matrix for electron-phonon coupling.

We expect multiphonon processes to play a role for all dielectric boundaries even for 
graphite boundaries, where the Debye energy is rather high, but not high enough to couple
the two lowest image states (see table~\ref{DebyeEnergy}). This coupling, on the other hand, 
is the rate-limiting one, that is, the one which determines the electron desorption time. 
Multiphonon processes remain important when in addition to image states also 
crystal-induced surface states or dangling bonds are included because these states,
being stronger bound than image states, are energetically deep in the gap and thus far away 
from the vacuum (plasma) level.

In our model we made the overall assumption that plasma electrons cannot enter the
plasma boundary (hard boundary condition at $z=0$). At least electrons with an 
energy larger than the projected energy gap of the solid, can however enter the plasma
boundary, scatter inside the material, before bouncing back to the surface, where they 
may be either re-emitted to the plasma or trapped in surface bound states. Processes 
inside the material can be thus only neglected when the projected energy gap is much 
larger than the typical energies of plasma electrons, that is, $E_g\gg k_BT_e$, and
when the floating potential $\bar{U}$ is approximately in the middle of this large gap. 

Here we come to a potentially very interesting point, in particular, as far as metallic 
plasma boundaries are concerned. According to Fig.~\ref{Copper} the projected energy
gap depends on the crystallographic orientation of the surface. Even planar metallic 
plasma boundaries will however almost never coincide with a single crystallographic plane. 
At best, they contain large, crystallographically well-defined facets, as we discussed in 
the context of spherical grains. Hence, the projected energy gap varies along the boundary. 
Regions can thus be expected where surface states are absent and plasma electrons can easily 
enter the boundary. In other regions large gaps prevent plasma electrons from entering 
the material. Instead they would sit in surface states. How all this affects the spatial
distribution of surface charges is an open question. 

For dielectric boundaries the projected energy gap is of the order of the intrinsic 
gap of the bulk material. It depends only weakly on the crystallographic plane. 
But a problem which concerns both metallic and dielectric surfaces is the existence
of crystal-induced surface states. We would expect them to be less important 
for physisorption of electrons. Being strongly bound and having a center of gravity 
very close to the surface or even inside the plasma boundary, the spatial overlap 
between unbound surface states and crystal-induced surface states should be 
rather small. Hence, the matrix element controlling sticking into or desorption
from a crystal-induced surface state should be much smaller than the corresponding 
matrix element involving weakly-bound polarization-induced surface states, which
are always exterior to the boundary and, on a microscopic scale, even relatively far
away from the surface. However, only a detailed study can show if our intuition is
correct.

Another problem concerning both metallic and dielectric plasma boundaries is surface
roughness. In our model, the plasma boundary is a well-defined mathematical plane. 
On the atomistic scale, and we actually do calculations on this scale, the surface is 
however not perfect. In a refined model for surface states this aspect, possibly in
conjunction with surface reconstructions~\footnote{Here we do not mean the 
reconstruction of the surface due to impacting plasma particles but the intrinsic 
reconstruction leading to geometrical differences between real terminations of 
crystals and ideal crystallographic planes~\cite{Spanjaard96}.} and chemical 
contamination has to be taken into account.

Throughout we implicitly assumed that bound surface states exist although the exact
surface potential supporting them is unknown. Since surface states have been detected
many 
times~\cite{DAG84,SH84,WHJ85,JDK86,EP90,EL94,Fauster94,EL95,HSR97,CSE99,Hoefer99,VPE07}
this assumption seems to be justified. Naturally, it would be desirable to calculate 
the surface potential from first principles. However, if not 
illusionary, it is at least very challenging, even when the plasma boundary is planar and 
crystallographically well defined. The quantum-mechanical exchange and correlation effects 
determining the tail of the surface potential are beyond the local-density approximation, 
the work-horse of most ab-initio packages for the calculation of the electronic structure 
of solids. Instead, non-local density functional theory~\cite{HFH98} has to be used which 
is much more complicated to implement. A compromise would be to calculate the potential
inside the boundary from an ab-initio local-density package and then continuously 
match this \textquotedblleft internal\textquotedblright~potential to the 
\textquotedblleft external\textquotedblright~potential deduced from a model Hamiltonian
of the type presented in appendix A. The model produces a diverging potential only in 
the simplest approximation. With methods adapted from bulk polaron theory~\cite{EM73,Barton81}
potentials could be deduced which are finite at the boundary and thus continuously 
matchable with the periodic crystal potential obtained from the local density approximation.

As in other branches of surface science~\cite{KG86,Spanjaard96}, a general strategy to 
short-circuit unknown microscopic details about the surface would be to work with simple, 
possibly analytically solvable models containing parameters that can be adjusted to 
experimentally measured quantities, for instance, the binding energy of surface states. 

For this strategy to work, experimental techniques suitable for directly probing the 
electronic properties of surfaces, for instance, inverse photoemission 
spectroscopy~\cite{DAG84,SH84,JDK86}, from which the binding energy and the lifetime of 
unoccupied electron surface states can be determined, have to be adapted to plasma 
boundaries. In addition, macroscopic quantities, such as the quasi-stationary surface 
charge, the surface temperature, and the temperature and density of the electrons in
the plasma have to be also known. So far, however, these combined data are not available
for any experiment. For sure, surface charges have been measured in dielectric barrier 
discharges~\cite{LLZ08,SLP07,SAB06,KCO04} and of course in complex plasmas, where in 
fact a great variety of techniques has been invoked to determine the charge of floating 
$\mu m$-sized dust particles~\cite{Ishihara07,FIK05,KRZ05,SV03,TAA00,TLA00,WHR95}. But 
particularly in the experiments measuring grain charges the diagnostics of the hosting 
plasma is usually missing. In addition, although it is possible to measure the temperature
of the grain~\cite{MBK08}, grain temperature and charge have not yet been measured 
simultaneously. For the microscopic modeling of surface charges it is however important
to know at least these two quantities.  

\section{Concluding remarks}

In this colloquium we proposed to treat the interaction of electrons and ions  
with inert plasma boundaries, that is, boundaries which stay intact during their 
exposure to the plasma, as a physisorption process involving surface states.
The sticking coefficients $s_{e,i}$ and desorption times $\tau_{e,i}$ can then be
calculated from microscopic models containing (i) a static potential supporting
bound and unbound surface states and (ii) a coupling of these states to an 
environment which triggers transitions between them. Microscopically, the sticking of 
an electron or ion to the surface corresponds then to a transition from an unbound 
surface state to a bound one. Desorption of an electron or ion from the wall is then 
simply the reverse process. 

Although this point of view can be applied to ions and electrons, we worked it out
-- for the particular case of a metallic boundary and within the simplest possible
model -- only for electrons because the surface states for electrons are surface 
states in the ordinary sense, that is, states which are only a few nanometers away from
the surface. The environment responsible for transitions between electron surface 
states, and thus for sticking and desorption of an electron, are therefore the elementary 
excitations of the solid. For ions, however, as soon as the surface collected some 
electrons, the surface potential is the long-range attractive Coulomb potential. 
Sticking and desorption of ions occurs thus far away from the surface. Nevertheless, 
provided the surrounding plasma is taken as the environment triggering transitions 
between ion surface states, the dynamics and kinetics of ions in front of the boundary 
can be described in close analogy to the electron dynamics and kinetics occurring much 
closer to the surface. Since without the surface no attractive Coulomb potential for 
ions would exist, the ion dynamics and kinetics is also a kind of surface physics although
it takes place far away from the surface. 

Ions are much heavier than electrons and the potential most relevant for them, the 
Coulomb potential, varies on a scale much larger than the ion's de-Broglie wavelength. 
Quantum mechanics is thus not really required for studying the ion kinetics in front
of a plasma boundary. Instead of pushing the quantum-mechanical techniques we used 
for electrons to the semiclassical regime, it is thus also possible to analyze ions 
with Boltzmann equations. In that case it is however crucial to set up two Boltzmann 
equations, one for unbound ions and one for bound ions. As in the quantum-mechanical
calculation, collisions of bound and unbound ions with the atoms/molecules of the background 
gas determine the number of trapped ions and how they are spatially distributed. 

Studying the ion dynamics and kinetics is important because it affects the rate with 
which ions and electrons may recombine in the vicinity of the grain surface. If the 
corresponding flux $\alpha_R\sigma_e\sigma_i$ is larger than the electron desorption 
flux $\tau_e^{-1}\sigma_e$, the charge of the grain is the one which balances on 
the grain surface the electron collection flux $s_ej_e^{\rm plasma}$ with the ion 
collection flux $s_ij_i^{\rm plasma}$ and not with the electron desorption flux
$\tau_e^{-1}\sigma_e$ as in our surface model for the grain charge. Provided 
$s_e\sim s_i$ this would eventually lead to the standard criterion from which the
grain charge is calculated. We emphasize in this respect however that the rate 
equations for the surface densities $\sigma_{e,i}$ are phenomenological. They 
should be derived from a quantum-mechanical surface scattering kernel taking 
bound surface states into account. Only then would we know if the microscopically
obtained $s_e$ and $\tau_e$ and the macroscopic plasma fluxes $j_{e,i}^{\rm plasma}$ 
are as simply connected as in the phenomenological rate equations. In any case, the 
quantum-mechanical approach for calculating $s_e$ and $\tau_e$ stands by itself 
irrespective of the fate of our surface model for the grain charge.

Admittingly, the microphysics at the plasma-boundary we discuss is
not the one utilized in plasma technology. Precisely the processes we excluded 
are most important there: Implantation of heavy particles, reconstruction or destruction
of the surface due to high energy particles, and chemical modification due to radicals, 
to name just a few. The target surfaces are of course charged but, from the perspective 
of plasma technology, the surface charges only control the particle fluxes to the 
surfaces. Properties other than their mere existence are of no concern. 

From a microscopic point of view, the technologically important surface processes just
listed are extremely complicated. A description of these processes at a level, let say, 
solid state physicists describe superconductivity in bulk metals is certainly far from reach. 
It may even not be required for plasma technology to proceed as a business. But as in other
branches of science, it is the pleasure and duty of research driven by curiosity to 
push the understanding of particular processes, technologically relevant or not, to an ever
increasing level of sophistication. We firmly believe, the microphysics at an inert plasma
boundary is now ready for a truly microscopic understanding.
It is our hope to have inspired other groups joining us on 
our journey to the microphysics at an inert plasma boundary. In particular, however, we 
hope to have found experimentalists eager to design experiments with well-defined model 
surfaces, which come as close as possible to the idealized boundaries theorists have to
consider in their calculations, and at the same time are accessible to the surface 
diagnostics used elsewhere in surface science.

\appendix

\section{Microscopic model for the image potential}

In this appendix we discuss a microscopic model which interprets the image potential
in terms of virtual excitation of surface modes~\cite{RM72,EM73,Barton81}. The model
is applicable to metals and dielectrics. 

To be specific, we consider a planar plasma boundary in the $xy$ plane putting the plasma in 
the positive halfspace defined by $z>0$. A convenient starting point for a microscopic description 
of the polarization-induced interaction between an electron and a boundary is the single electron
Hamiltonian~\cite{RM72,EM73},
\begin{eqnarray}
H&=&-\frac{\hbar^2}{2m_e}\Delta+\hbar\omega_s\sum_{\vec{K}}a_{\vec{K}}^\dagger a_{\vec{K}}
\nonumber\\
&+&\sum_{\vec{K}}\Gamma(K)\exp[-i\vec{K}\cdot\vec{R}-Kz](a^\dagger_{\vec{K}}+a_{\vec{-K}})~,
\label{StartingH}
\end{eqnarray}
where $\Delta$ is the three-dimensional Laplace operator, $a^\dagger_{\vec{K}}$ is the
creation operator for the polarization-induced surface mode responsible for the interaction,
and
\begin{eqnarray}
\Gamma(K)=\bigg(\frac{\pi e^2\hbar\omega_s}{AK}\cdot\frac{\epsilon-1}{\epsilon+1}\bigg)^{\frac{1}{2}}
\end{eqnarray}
is the coupling function; $\vec{K}=(K_x,K_y)$ is a two-dimensional wavevector, $\vec{R}=(x,y)$ 
denotes the projection of the electron position onto the surface, whose area is $A$, and $z$ 
is the distance of the electron from the surface. 

For metals ($\epsilon=\infty$), the relevant surface
modes are surface plasmons with typical energies of a few electron volts, for instance, for 
copper, $\hbar\omega_s\approx 2eV$~\cite{EL95}. For dielectrics ($\epsilon < \infty$), the 
relevant surface modes are optical phonons with energies of a few tenth of an electron volt, 
for instance, for graphite, $\hbar\omega_s\approx 0.43eV$ when we use
$\omega_s=\omega_T\sqrt{(\epsilon+1)/2}$ with $\epsilon=12$ and $\omega_T=0.17eV$~\cite{MRT04}.

To approximately separate the static from the dynamic interaction, we apply to the
Hamiltonian~(\ref{StartingH}) the unitary transformation~\cite{Barton81}
\begin{eqnarray}
U=\exp\bigg[\sum_{\vec{K}}\big(\gamma^*_{\vec{K}}(\vec{R},z)a^\dagger_{\vec{K}}-
\gamma_{\vec{K}}(\vec{R},z)a_{\vec{K}}\big)\bigg]
\end{eqnarray}
with
\begin{eqnarray}
\gamma_{\vec{K}}(\vec{R},z)=\frac{\Gamma(K)}{\hbar\omega_s}\exp[i\vec{K}\cdot\vec{R}-Kz]~.
\end{eqnarray}
After the transformation the Hamiltonian reads 
\begin{eqnarray}
\bar{H}&=&UHU^\dagger
\nonumber\\
&=&-\frac{\hbar^2}{2m_e}\Delta+V_p(z)+
\hbar\omega_s\sum_{\vec{K}}a_{\vec{K}}^\dagger a_{\vec{K}}
\nonumber\\
&+&\frac{i\hbar}{m_e}\vec{A}(\vec{r})\cdot\nabla + \frac{1}{2m_e}\vec{A}(\vec{r})\cdot\vec{A}(\vec{r})
\label{Hbar}
\end{eqnarray}
where 
\begin{eqnarray}
V_p(z)&=&
-\sum_{\vec{Q}}\hbar\omega_s|\gamma_{\vec{Q}}(\vec{R},z)|^2
\nonumber\\
&=&-\frac{e^2(\epsilon-1)}{4(\epsilon+1)z}
\label{classVp}
\end{eqnarray}
is the classical image potential arising from virtual excitation of surface 
modes and
\begin{eqnarray}
\vec{A}(\vec{r})=-i\hbar\sum_{\vec{K}}\bigg[\nabla\gamma^*_{\vec{K}} a_{\vec{K}}^\dagger - h.c.\bigg]
\end{eqnarray}
is a vector potential giving rise to a \textquotedblleft minimal-type\textquotedblright~
dynamic coupling-- the last two terms on the rhs of~(\ref{Hbar})-- 
between the electron and the surface mode. 

The first two terms on the rhs of~(\ref{Hbar}) describe an electron in a potential. Diagonalizing
these two terms, that is, using the eigenstates of Eq.~(\ref{Schroedinger3D}) with 
$V(z)\rightarrow V_p(z)$ as a basis and ignoring the nonlinear term $\sim \vec{A}^2$ we obtain 
\begin{eqnarray}
\bar{H}&=&H_e
+\hbar\omega_s\sum_{\vec{K}}a_{\vec{K}}^\dagger a_{\vec{K}}
\nonumber\\
&+&\!\!\sum_{\vec{Q},\vec{K}}\sum_{q,q'}G_{qq'}(\vec{Q},\vec{K})(a_{\vec{K}}^\dagger-a_{-\vec{K}})
C_{\vec{Q}-\vec{K}q}^\dagger C_{\vec{Q}q'}~,
\label{TransformedH}
\end{eqnarray}
where 
\begin{eqnarray}
H_e=\sum_{\vec{Q}q}E_{\vec{Q}q}C_{\vec{Q}q}^\dagger C_{\vec{Q}q}
\end{eqnarray}
describes an electron in classical image states. Thus, without the dynamic coupling to surface
modes, $\bar{H}\rightarrow H_e$, and we would have obtained the model we used for the 
calculation of $s_e$ and $\tau_e$. 

Obviously, the dynamic coupling encoded in the last term on the rhs of~(\ref{TransformedH})
renormalizes the classical image states. The eigenstates of the full Hamiltonian -- the 
true polarization-induced surface states -- are not identical to the classical image states.
The latter should be considered as zeroth order (or bare) eigenstates. Better approximations 
can be constructed using methods from polaron theory~\cite{EM73,Barton81}. At large enough 
distances, however, where the residual interaction becomes negligibly small, classical image 
states are reasonably good approximations to the true polarization-induced surface states. 

Separating the lateral from the vertical motion according to Eqs.~(\ref{Schroedinger3D})
and~(\ref{SGpsi}) the matrix element for the dynamic coupling between the electron and the surface 
mode becomes
\begin{eqnarray}
G_{qq'}(\vec{Q},\vec{K})=\frac{\hbar\Gamma(K)}{m\omega_s A}
\bigg[\vec{Q}\!\cdot\!\vec{K}J^{(1)}_{qq'}(K)\!-\!KJ^{(2)}_{qq'}(K)\bigg]
\label{Gmatrix}
\end{eqnarray}
with 
\begin{eqnarray}
J^{(1)}_{qq'}(K)&=&\int dz \psi^*_q(z)\exp[-Kz]\psi_{q'}(z)~,
\label{J(1)}\\
J^{(2)}_{qq'}(K)&=&\int dz \psi^*_q(z)\exp[-Kz]\frac{d}{dz}\psi_{q'}(z)~.
\label{J(2)}
\end{eqnarray}

In general, $G_{qq'}$ is non-diagonal. It contains intraband ($q=q'$) and interband 
($q\neq q'$) transitions. The latter could in principle affect the physisorption kinetics
of electrons (understood -- for the moment -- as transitions between bound and unbound bare 
surface states). This happens however only when the energy of the surface mode is comparable 
to $k_BT_s$, where $T_s$ is the surface temperature, as well as comparable to the energy spacing 
of the bare surface states. Transitions between bare surface states are then associated with
creating or annihilating real surface modes in contrast to virtual modes which would only
renormalize the energies $E_{\vec{Q}q}$ and the wavefunctions $\psi_{\vec{Q}q}$. 
Physisorption would then be triggered by other elementary excitations of the solid, for 
instance, phonons and would moreover take place between renormalized surface states.

Whether the residual interaction directly triggers physisorption of electrons or not depends
on the material. For dielectric boundaries, for instance, graphite, 
$\hbar\omega_s \sim\Delta E_{21}$ (see table~\ref{DebyeEnergy}).  The nondiagonal elements
of $G_{qq'}$ could thus indeed be important for the physisorption process, especially at 
high temperatures. For metallic boundaries, however, the energy of the surface plasmon is a
few electron volts and thus far too high to play any direct role in the physisorption process. 
The dynamic coupling to surface plasmons modifies then primarily the properties of the surface 
states in which physisorption takes place (energy, wavefunction). We neglected these modifications, 
although at short distances they are not necessarily small on the scale of the electron binding 
energy, which is the relevant energy scale. We thereby assumed that for all distances, not only 
for large distances, the true polarization-induced surface states can be reasonably well 
approximated by classical image states.

\onecolumn

\section{Electronic wavefunctions and matrix elements}

In this appendix we summarize the properties of the electronic wavefunctions for the vertical
motion of the external electron. The results are well-known and the appendix primarily serves
the purpose to fix our notation.

First, we consider bound surface states. Using $y=z/2a_Bn$, with $n=1,2,...$ the quantum number
labelling the Rydberg series of bound states, the Schr\"odinger equation~(\ref{SGpsi})
for the vertical motion becomes 
\begin{eqnarray}
\frac{d^2}{dy^2}\psi_n(y) +\bigg[-\frac{1}{4}+\frac{n}{y}\bigg]\psi_n(y)=0~,
\label{Whitteq}
\end{eqnarray}
whose solutions are Whittaker functions~\cite{AS73}. Hence, the wavefunctions which vanish 
at $z=0$ and for $z\rightarrow\infty$ are 
\begin{eqnarray} 
\psi_n(z)=N_n W_{n,1/2}(y)=\exp[-y/2]y(-)^{n-1}(n-1)!L_{n-1}^{(1)}(y)~, 
\end{eqnarray} 
where $N_n$ is a normalization constant and $L_{n-1}^{(1)}(y)$ is an associated Laguerre 
polynomial~\cite{AS73}. The corresponding eigenvalues are $E_n=-R_0/16n^2$.

In order to find the normalization constant, we insert the expansion of the Whittaker 
function~\cite{AS73}, 
\begin{eqnarray} 
W_{n,1/2}(y)&=&\exp[-y/2]y^n\sum_{q=0}^n a_q y^{-q}~, 
\label{WnExp} 
\end{eqnarray}
where
\begin{eqnarray}
a_q&=&\frac{(-)^q}{q!}
     \frac{\Gamma(n+1)\Gamma(n)}{\Gamma(n-q)\Gamma(n-q+1)}
\end{eqnarray} 
with $\Gamma(n)$ the Gamma function, in the normalization integral,
\begin{eqnarray}
1=\int_0^\infty dz |\psi_n(z)|^2~.
\end{eqnarray} 
Term-by-term integration  leads then to 
\begin{eqnarray}
N_n=\sqrt{\frac{1}{4n^3\Gamma(n)^2a_B}}=\frac{{\cal N}_n}{\sqrt{a_B}}~,
\end{eqnarray}
which is the defining equation for ${\cal N}_n$ needed in appendix C.

For the continuum states, we use $y=ikz/2a_B$ as an independent variable. The 
Schr\"odinger equation~(\ref{SGpsi}) can then be reduced to~(\ref{Whitteq})
with $n$ replaced by $-ik^{-1}$. The continuum states with energy 
$E_k=R_0k^2/16$ which vanish at $z=0$ are 
thus given by~\cite{AS73} 
\begin{eqnarray}
\psi_k(z)=N_k M_{-ik^{-1},1/2}(y)~.
\end{eqnarray}

As for any continuum state, to find the normalization constant $N_k$ is somewhat
tricky. We could normalize $\psi_k(z)$ on the momentum scale but we found it 
more convenient to use a box-normalization considering the plasma halfspace 
($z>0$) as a slap of width $L$ with $L\rightarrow\infty$ at the end of the 
calculation. Thus, $N_k$ is determined from the 
condition
\begin{eqnarray}
1=\int_0^L dz |\psi_k(z)|^2~.
\label{NormIntk}
\end{eqnarray}

To do the normalization integral, we utilize the fact that in the limit $L\rightarrow\infty$ 
the contribution to the integral coming from small $z$ is negligibly small compared to the 
contribution coming from large $z$. Hence, we can replace in~(\ref{NormIntk}) 
$\psi_k(z)$ by its asymptotic form for large $z$:
\begin{eqnarray}
\psi_k(z)&\sim&\psi^{\rm in}_k(z) + \psi^{\rm out}_k(z) 
\\
&=& N_k\bigg[ \frac{\exp[-\pi/2k]}{\Gamma(1+ik^{-1})}\exp[ikx/4]
+\frac{\exp[-\pi/2k+i\pi]}{\Gamma(1-ik^{-1})}\exp[-ikx/4] \bigg]~,
\label{AsymptExp}
\end{eqnarray}
where we defined in- and outgoing waves which we need in appendix C for the calculation 
of $s_e$.
 
The normalization constant is then given by 
\begin{eqnarray}
N_k=\sqrt{\frac{\pi}{Lk(1-\exp[-2\pi/k])}}=\frac{{\cal N}_k}{\sqrt{L}}
\end{eqnarray}
which also defines ${\cal N}_k$ needed in appendix C.

Having appropriately normalized wavefunctions, we can now calculate the electronic
matrix element~(\ref{I(1)}). Although we could calculate~(\ref{I(1)}) for any 
$\vec{R}$ and any $k$ we give only the result for $\vec{R}=0$ and $k\ll 1$ because 
in the calculation of $s_e$ and $\tau_e$ we eventually approximate (\ref{I(1)}) by 
$I^{(1)}_{nk\ll 1}(0)$. The  multidimensional integrals defining $s_e$ and $\tau_e$ 
are then easier to perform. 

The matrix element we need is 
\begin{eqnarray}
I^{(1)}_{nk}(0)=2n{\cal N}_n{\cal N}_k\int_0^\infty dy \exp[-y/d]W_{n,1/2}(y)M_{-ik^{-1},1/2}(ikny)~.
\label{Inq1}
\end{eqnarray}
Approximating ${\cal N}_k\approx (\pi/k)^{1/2}$ for $k\ll 1$ gives 
\begin{eqnarray}
I^{(1)}_{nk\ll 1}(0)=\sqrt{\frac{\pi}{nk}}\frac{1}{\Gamma(n)} I_1
\end{eqnarray}
with 
\begin{eqnarray}
I_1=\int_0^\infty dy \exp[-y/d]W_{n,1/2}(y)M_{-ik^{-1},1/2}(ikny)~,
\end{eqnarray}
which, to be consistent, we also have to calculate for $k\ll 1$.

To determine the integral $I_1$, we use the expansion~(\ref{WnExp}) for $W_{n,1/2}(y)$ 
together with the expansion~\cite{AS73} 
\begin{eqnarray}
M_{-ik^{-1},1/2}(ikny)=ikn\sum_{m=0}^\infty C_m \frac{(ikn)^m}{[n(1-ik)]^{(m+1)/2}}
y^{(m+1)/2}J_{m+1}(2\sqrt{ny(1-ik)})
\end{eqnarray}
for $M_{-ik^{-1},1/2}(ikny)$, where $C_m$ are constants and $J_n(y)$ are 
Bessel functions~\footnote{Specifically, we employ formula 13.3.8 from~\cite{AS73} with $h=1/2$.}. 
Thus,
\begin{eqnarray}
I_1=ikn\sum_{m=0}^\infty\sum_{q=0}^n a_q C_m \frac{(ikn)^m}{[n(1-ik)]^{(m+1)/2}} I_2
\label{Inte1}
\end{eqnarray}
with an integral $I_2$ which can be found in~\cite{Gradstein81}:
\begin{eqnarray}
I_2&=&\int_0^\infty dy \exp[-(1/d+1/2)y]\exp[n-q+(m+1)/2]J_{m+1}(2\sqrt{ny(1-ik)})
\label{Inte2}
\nonumber\\
&=&\bigg(\frac{2d}{2+d}\bigg)^{n-q+m+2}(n-q)![(1-ik)n]^{(m+1)/2}
\exp\bigg[-(1-ik)\frac{2nd}{2+d}\bigg]L_{n-q}^{(m+1)}((1-ik)\frac{2nd}{2+d})
\end{eqnarray}
with $d=1/2nk_s$. 

Inserting~(\ref{Inte2}) for $k\ll 1$ into~(\ref{Inte1}) and using $C_0=1$~\cite{AS73} we 
finally obtain
\begin{eqnarray}
\big|I^{(1)}_{nk\ll 1}(0)\big|^2=k\big|I^{(1)}_n\big|^2=\pi n k \bigg(\frac{2d}{2+d}\bigg)^4|f_n|^2
\label{In1}
\end{eqnarray}
with 
\begin{eqnarray}
|f_n|^2=\sum_{q=0}^n\frac{(-)^q}{q!}\frac{\Gamma(n+1)}{\Gamma(n-q)}\bigg(\frac{2d}{2+d}\bigg)^{n-q}
\exp\bigg[-\frac{2nd}{2+d}\bigg]L_{n-1}^{(1)}(\frac{2nd}{2+d})~, 
\end{eqnarray}
where Eq.~(\ref{In1}) defines $|I^{(1)}_n|^2$ used in appendix C.

Finally we give the result for the matrix element~(\ref{I(2)}) for $\vec{R}=0$. Using the single 
electron states of the metal specified in~(\ref{StatesForInternalElectron}) and measuring length
again in units of $a_B$, 
\begin{eqnarray}
I^{(2)}_{kk'}(0)&=&2\int_0^\infty dx \exp[-k_sx]sin(kx)sin(k'x) 
\label{Ikk2}\\
&=&16k_s^2\frac{\sqrt{E_kE_{k'}}}
{[(k_s^2+E_{k}+E_{k'})^2-4E_kE_{k'}]^2} kk'
\\
&=&16k_s^2J^{(2)}(E_k,E_{k'})kk'~,
\end{eqnarray}
which also defines the function $J^{(2)}(E_k,E_{k'})$, with $E_k=k^2$
and likewise for $E_{k'}$, needed in appendix C.  

\section{Calculation of $s_e$ and $\tau_e$}

In this appendix we give mathematical details concerning the calculation of $s_e$ and $\tau_e$. In 
all equations below we use dimensionless variables measuring energies and lengths in units of $R_0$
and $a_B$, respectively. We are furthermore interested in the limit $L\rightarrow\infty$ and 
$A\rightarrow\infty$. Thus, momentum sums become integrals according to
\begin{eqnarray}
\frac{1}{L}\sum_k=\int \frac{dk}{2\pi}~~~~~~\mbox{and}~~~~~~~
\frac{1}{A}\sum_{\vec Q}&=&\int \frac{d\vec{Q}}{(2\pi)^2}~. 
\end{eqnarray}

For the purpose of doing some of the integrals analytically, we found it convenient to 
rewrite the $\delta-$function for energy conservation as follows:
\begin{eqnarray}
\delta(E_{\vec{Q}'q'}-E_{\vec{Q}n}+E_{\vec{K}'k'}-E_{\vec{K}k})=
\int_{-\infty}^\infty d\omega \delta(E_{\vec{Q}'q'}-E_{\vec{Q}n}-\omega)
\delta(E_{\vec{K}'k'}-E_{\vec{K}k}+\omega)~.
\label{auxE}
\end{eqnarray}
The angles can then be integrated out and the global sticking coefficient $s_e$ defined 
in Eq.~(\ref{seSum}) becomes
\begin{eqnarray}
s_e&=&\frac{4\beta_e^{3/2}}{\pi^2\beta_s^{1/2}}
\sum_n\int_0^\infty dq'\int_0^\infty dk\int_0^\infty dk'
\int_{-\infty}^\infty d\omega\int_0^\infty dR
~\frac{\big|I_{nq'}^{(1)}(0)I_{kk'}^{(2)}(0)\big|^2}{k_s^2+R^2}
\nonumber\\
&\times&[1+n_B(\omega)]N(R,\omega,E_k,E_{k'})R^{-1}\exp[-\beta_e\Psi_n(R,\omega,E_{q'})]~,
\end{eqnarray}
where, for simplicity, we have neglect the dependence of the electronic matrix 
elements~(\ref{I(1)}) and~(\ref{I(2)}) on the lateral momentum transfer $\vec{R}=\vec{Q}-\vec{Q}'$ 
and introduced two functions:
\begin{eqnarray}
N(R,\omega,E_k,E_{k'})&=&
F_{-1/2}(\beta_s(E_F-y_{kk'}(R,\omega)+\omega)-F_{-1/2}(\beta_s(E_F-y_{kk'}(R,\omega)))~,
\\
\Psi_n(R,\omega,E_{q'})&=&E_n+\omega+
\bigg(\frac{E_{q'}-E_n+R^2-\omega}{2R}\bigg)^2~,
\end{eqnarray}
with $F_{-1/2}(x)$ Fermi integrals for which, as far as the numerics is concerned, we take Unger's 
approximation~\cite{Unger88}, and
\begin{eqnarray}
y_{kk'}(R,\omega)=E_k+\bigg(\frac{E_k-E_{k'}-R^2-\omega}{2R}\bigg)^2
\label{ykk}
\end{eqnarray}
with $E_k=k^2$ and likewise for $E_k$ and $E_{q'}$. The functions 
$I_{nq}^{(1)}(0)$ and $I_{kk'}^{(2)}(0)$ are, respectively, defined in Eqs.~(\ref{Inq1}) 
and~(\ref{Ikk2}) in appendix B.

Using $E_{q'}$, $E_k$, and $E_{k'}$ instead of $q'$, $k$, and $k'$ as integration variables, we 
finally find the result presented in Eqs.~(\ref{seFinal}) and~(\ref{Istick}) with 
\begin{eqnarray}
h(R,\omega)&=&\sum_n\big|I_n^{(1)}\big|^2\exp[-\beta_e(E_n+\omega)]
\int_{x_n(R,\omega)}^\infty dx \exp[-\beta_e x^2]~,
\label{hfct}\\
g(R,\omega)&=&\int_0^\infty dE dE' J^{(2)}(E,E')N(R,\omega,E,E')~,
\label{gfct}
\end{eqnarray}
and 
\begin{eqnarray}
x_n(R,\omega)=\frac{R^2-E_n-\omega}{2R}~.
\end{eqnarray}

The calculation of the energy resolved sticking coefficient proceeds along the same lines. For 
perpendicular incidence we find the result stated in Eq.~(\ref{sperp}) of the main text with 
\begin{eqnarray}
g^\perp(R,E')=\sum_n\big|I^{(1)}_n\big|^2 \frac{1+n_B(E'-E_n-R^2)}
{1+(R/k_s)^2} g(R,E'-E_n-R^2)~.
\label{gperp}
\end{eqnarray}

Now we turn our attention to the calculation of the desorption time. It is quite similar. An 
intermediate expression, after expressing energy conservation in the form~(\ref{auxE}) and
performing the integrals over angles, is
\begin{eqnarray}
\tau_e^{-1}&=&\frac{R_0}{8\pi^3\hbar Z}
\sum_{n'}\int_0^\infty dq\int_0^\infty dk\int_0^\infty dk'
\int_{-\infty}^\infty d\omega\int_0^\infty dR
~\frac{\big|I_{qn'}^{(1)}(0)I_{kk'}^{(2)}(0)\big|^2}{k_s^2+R^2}
\nonumber\\
&\times&[1+n_B(\omega)]N(R,\omega,E_k,E_{k'})R^{-1}\exp[-\beta_s\Phi_{n'}(R,\omega,E_{q})]~,
\end{eqnarray}
where we have again neglected the dependence of the electronic matrix elements 
~(\ref{I(1)}) and~(\ref{I(2)}) on the transfer of lateral momentum and introduced
\begin{eqnarray}
Z&=&\sum_n\exp[-\beta_sE_n]~,
\\
\Phi_{n'}(R,\omega,E_q)&=&E_{q}+\omega+
\bigg(\frac{E_{q}-E_{n'}-R^2+\omega}{2R}\bigg)^2~.
\end{eqnarray}

Using again $E_k=k^2, E_{k'}=k'^2$, and $E_q=q^2$ as integration variables we 
finally find the result~(\ref{taue}) and~(\ref{Idesorb}) given in main text with 
\begin{eqnarray}
f(R,\omega)=\sum_n\big|I_n^{(1)}\big|^2
\exp[-\beta_eE_n]
\int_{y_n(R,\omega)}^\infty dy \exp[-\beta_s y^2]
\label{ffct}
\end{eqnarray}
and
\begin{eqnarray}
y_n(R,\omega)=\frac{\omega+R^2-E_n}{2R}~.
\end{eqnarray}

At the end of this appendix let us say a few words about Laplace's approximation~\cite{Olver74} 
which we used to derive Eqs.~(\ref{seL})--(\ref{stauL}). If there is only a 
single bound state with energy $E_1$ the summations over $n$ reduce to a single term. 
For $k_BT_s\ll |E_1|$ and $k_BT_e\ll |E_1|$ it is then possible to do some of the integrals defining 
$s_e$ and $\tau_e$ asymptotically within Laplace's approximation.

First, we consider Laplace's approximation for $\tau_e$. For a single bound state  
\begin{eqnarray}
f(R,\omega)=\big|I_1^{(1)}\big|^2\exp[-\beta_eE_1]\int_{y_1(R,\omega)}^\infty dy \exp[-\beta_s y^2]~.
\end{eqnarray}
Provided $k_BT_s\ll|E_1|$, $f(R,\omega)$ is largest for $y_1(R,\omega)\le 0$, that is, for 
$\omega\le E_1-R^2<0$. In this domain, Laplace's approximation to the $y-$integral gives 
$\sqrt{\pi/\beta_s}/2$, where the factor $1/2$ anticipates that the $R$- and 
$\omega$-integrations are later performed also within Laplace's approximation. Changing 
$\omega\rightarrow -\omega$, we obtain
\begin{eqnarray}
I_{\rm desorb}^L\approx\frac{1}{2}\frac{\pi^{1/2}}{\beta_s^{1/2}}\int_0^\infty dR\int_{|E_1|+R^2}^\infty d\omega
\frac{n_B(\omega)}{1+(R/k_s)^2}
\big|I_1^{(1)}\big|^2\exp[-\beta_eE_1][-g(R,-\omega)]~,
\end{eqnarray}
where we used $1+n_B(-\omega)=-n_B(\omega)$. Hence, using~(\ref{taue})
\begin{eqnarray}
\tau_e^L\approx\frac{4\pi^{5/2}\beta_s^{1/2}\hbar}{R_0\big|I_1^{(1)}\big|^2 J(|E_1|)}~,
\label{auxTaue}
\end{eqnarray}
where 
\begin{eqnarray}
J(|E_1|)=\int_0^\infty dR\int_{|E_1|+R^2}^\infty d\omega\frac{n_B(\omega)}{1+(R/k_s)^2}
[-g(R,-\omega)]~.
\label{Jfct}
\end{eqnarray}
Since $k_BT_s\ll |E_1|$, we can approximate in~Eq.~(\ref{Jfct}) the Bose distribution function 
$n_B(\omega)$ by $\exp[-\beta_s\omega]$. Hence, the main contribution to the $\omega-$integral will 
come from its lower boundary. Calculating the $\omega-$integral within Laplace's approximation 
and then applying, in a last step, Laplace's approximation also to the remaining $R-$integral,
we find
\begin{eqnarray}
J(|E_1|)\approx\frac{\pi^{1/2}}{2\beta_s^{3/2}}\bar{g}\exp[-\beta_s|E_1|]
\end{eqnarray}
with 
\begin{eqnarray}
\bar{g}=\lim_{R\rightarrow 0}[-g(R,-|E_1|)]~,
\label{gbar}
\end{eqnarray}
which, combined with Eq.~(\ref{auxTaue}), leads to Eq.~(\ref{taueL}) given in the main text.

Calculating $s_e$ within Laplace's approximation is quite similar. However, whereas for $\tau_e$
it is a reasonable approximation, because $k_BT_s\ll |E_1|$, for $s_e$ it is only meaningful 
when $k_BT_e$ is also much smaller than $|E_1|$. Under this assumption, which is of course 
usually not satisfied, we find from Eq.~(\ref{Istick}), again anticipating that the 
$R$- and $\omega$-integrations are later performed within Laplace's approximation, 
\begin{eqnarray}
I^L_{\rm stick}\approx\frac{1}{2}\frac{\pi^{1/2}}{\beta_e^{1/2}}\big|I^{(1)}_1\big|^2\exp[\beta_e|E_1|]K(|E_1|)
\label{IstickL}
\end{eqnarray}
with
\begin{eqnarray}
K(|E_1|)=\int_0^\infty dR\int_{|E_1|+R^2}^\infty d\omega
\frac{1+n_B(\omega)}{1+(R/k_s)^2}
\exp[-\beta_e\omega]g(R,\omega)~,
\end{eqnarray}
to which we again successively apply Laplace's approximation to find
\begin{eqnarray}
K(|E_1|)\approx\frac{\pi^{1/2}}{2\beta_e^{3/2}}\tilde{g}\exp[-\beta_e|E_1|] 
\label{Kfct}
\end{eqnarray}
with
\begin{eqnarray}
\tilde{g}=\lim_{R\rightarrow 0}[g(R,|E_1|)]~.
\label{gtilde}
\end{eqnarray}
Hence, combining~(\ref{Kfct}) and~(\ref{IstickL}) and inserting the result in~(\ref{seFinal}) 
gives
\begin{eqnarray}
s_e^L=\frac{4\big|I^{(1)}_1\big|^2\tilde{g}}{\pi \beta_s^{1/2}\beta_e^{1/2}}~.
\label{seLaux}
\end{eqnarray}

Using the properties of the function $g(R,\omega)$ we now show that $\tilde{g}=\bar{g}$. First,
we see from the definition~(\ref{gfct}) that the $R-$dependence of $g(R,|E_1|)$ comes 
from the $R-$dependence of the function $y_{kk'}(R,|E_1|)$ defined in~(\ref{ykk}). Then we 
notice that 
\begin{eqnarray}
\lim_{R\rightarrow 0}y_{kk'}(R,|E_1|)=
\left\{ \begin{array}{cc}
         \infty   & {\rm for~E\neq E'+|E_1|}\\
         & \\
         E & {\rm for~E=E'+|E_1|}
       \end{array}\right. ~,
\end{eqnarray}
from which follows
\begin{eqnarray}
\lim_{R\rightarrow 0}N(R,|E_1|,E,E')=
\left\{\begin{array}{cc}
0   & {\rm for~E\neq E'+|E_1|}\\
& \\
F_{-1/2}(\beta_s(E_F-E+|E_1|))-F_{-1/2}(\beta_s(E_F-E)) & {\rm for~E=E'+|E_1|}
\end{array}\right.~,
\end{eqnarray}
because $F_{-1/2}(x)$ vanishes for $x\rightarrow -\infty$, and thus 
\begin{eqnarray}
\tilde{g}&=&\lim_{R\rightarrow 0}g(R,|E_1|)
\\
&=&\int_{0}^\infty dE' J^{(2)}(E'+|E_1|,E')
\big[F_{-1/2}(\beta_s(E_F-E'))
-F_{-1/2}(\beta_s(E_F-E'-|E_1|))\big]
\\
&=&\int_{0}^\infty dE' J^{(2)}(E',E'+|E_1|)
\big[F_{-1/2}(\beta_s(E_F-E'))
-F_{-1/2}(\beta_s(E_F-E'-|E_1|))\big]~,
\end{eqnarray}
where in the last line we used $J^{(2)}(E,E')=J^{(2)}(E',E)$. To calculate $\bar{g}$ we 
proceed in the same way, noticing however that $N(R,-|E_1|,E,E')$ is finite only for
$E'=E+|E_1|$. Hence, 
\begin{eqnarray}
\bar{g}&=&\lim_{R\rightarrow 0}[-g(R,-|E_1|)]
\\
&=&-\int_{0}^\infty dE J^{(2)}(E,E+|E_1|)
\big[F_{-1/2}(\beta_s(E_F-E-|E_1|))
-F_{-1/2}(\beta_s(E_F-E))\big]
\\
&=&\int_{0}^\infty dE J^{(2)}(E,E+|E_1|)
\big[F_{-1/2}(\beta_s(E_F-E))
-F_{-1/2}(\beta_s(E_F-E-|E_1|))\big]
\\
&=&\tilde{g}~.
\end{eqnarray}
Since $\tilde{g}=\bar{g}$, Eq.~(\ref{seLaux}) is identical to Eq.~(\ref{seL}) given in 
the main text. 

\twocolumn

\begin{acknowledgement}
Support from the SFB-TR 24 ``Complex Plasmas'' and discussions with 
H. Kersten are greatly acknowledged. In the early stages of this 
work F.~X.~B. was funded by MV 0770/461.01. He is also grateful to M. 
Lampe for a particularly illuminating discussion.
\end{acknowledgement}

%
% BibTeX users please use
\bibliographystyle{Style/icpig}
\bibliography{MicroPhys}

\end{document}